\documentclass[journal]{elsarticle}
\usepackage[paperwidth=18.4cm,paperheight=26cm,top=1.5cm,bottom=2cm,right=2cm]{geometry}
\usepackage{amssymb}
\usepackage{amsmath}
\usepackage{hyperref} 
\hypersetup{
colorlinks=true,
linkcolor=blue,
filecolor=blue,
urlcolor==blue,
citecolor=blue,
}
\biboptions{sort&compress} 
\usepackage{color}
\usepackage{setspace}
\usepackage[noend]{algpseudocode}
\usepackage{graphicx}
\usepackage{float} 
\usepackage{subfigure}
\usepackage{algorithmicx,algorithm}
\usepackage{multirow}
\usepackage{bbm}
\def\degree{${}^{\circ}$}

\journal{Aerospace Science and Technology (AESCTE) }

\begin{document}

\begin{frontmatter}

\title{Fast sparse flow field prediction around airfoils via multi-head perceptron based deep learning architecture}

\cortext[mycorrespondingauthor]{Corresponding author}

\author[nwpu_address,International_institute,cardc_address]{Kuijun Zuo}

\author[nwpu_address]{Shuhui Bu}

\author[nwpu_address,International_institute]{Weiwei Zhang \corref{mycorrespondingauthor}}
\ead{aeroelastic@nwpu.edu.cn}

\author[nwpu_address,International_institute]{Jiawei Hu}

\author[nwpu_address]{Zhengyin Ye}

\author[cardc_address]{Xianxu Yuan}

\address[nwpu_address]{School of Aeronautics, Northwestern Polytechnical University, Xi'an, 710072, China}
\address[International_institute]{International Joint Institute of Artificial Intelligence on Fluid Mechanics, Northwestern Polytechnical University, Xi'an, 710072, China}
\address[cardc_address]{State Key Laboratory of Aerodynamics, China Aerodynamics Research and Development Center, Mian'yang, Si'chuan 621000, China}

\begin{abstract}
{ \indent 
In order to obtain the information about flow field, traditional computational fluid dynamics methods need to solve the Navier-Stokes equations on the mesh with boundary conditions, which is a time-consuming task.
In this work, a data-driven method based on convolutional neural network and multi-head perceptron is used to predict the incompressible laminar steady sparse flow field around the airfoils.
Firstly, we use convolutional neural network to extract the geometry parameters of the airfoil from the input gray scale image. 
Secondly, the extracted geometric parameters together with Reynolds number, angle of attack and flow field coordinates are used as the input of the multi-layer perceptron and the multi-head perceptron. 
The proposed multi-head neural network architecture can predict the aerodynamic coefficients of the airfoil in seconds.
Furthermore, the experimental results show that for sparse flow field, multi-head perceptron can achieve better prediction results than multi-layer perceptron.
}
\end{abstract}


\end{frontmatter}


\section{Introduction}

The design of the airfoil is a long-term development process \cite{DURU2022105312}. 
To meet the needs of different scenarios, a variety of airfoil families have been developed for different flight tasks, such as low-speed flight missions generally use front circular and posterior pointed airfoils, front and rear pointed airfoils are generally used for high-speed missions.
With the development of airfoil design, many numerical representation methods of airfoil geometry have been designed \cite{Hicks, BOEHM198717, Sobieczky1999, Stephen9269}, such as the class function/shape function transformation (CST) method using a shape function and a class function to describe an airfoil.
In the early stages of aircraft design, the appropriate airfoil is usually selected from the existing airfoil family.
With the development of machine learning technology, some scholars also try to apply machine learning method to airfoil design optimization and aerodynamic shape optimization \cite{du2021rapid, li2021deep, wang2021dual, li2022low, wu2022missile}.
As an important reference index of the airfoil, geometric parameters, Reynolds number and angle of attack are closely related to the aerodynamic performance of airfoils.
In particular, obtaining the aerodynamic parameters around the airfoils is critical to the airfoil selection and aircraft design.

In order to obtain the aerodynamic coefficients around the airfoils, such as lift, resistance, pressure, velocity, etc., wind tunnel test and computational fluid dynamics (CFD) techniques are two common methods. 
For wind tunnel testing, the design of the experimental process generally relies on the prior knowledge of experts, and the wind tunnel test process is not only long time required but also expensive.
Wind tunnel test is mainly used in the later stage of aircraft wing design, for comprehensive and accurate evaluation of the aerodynamic performance of aircraft wings. 
With the rapid development of computer hardware, the computing speed and performance of CFD have also been greatly improved.  
However, in terms of aerodynamics related design, many iterative processes need to be involved in the calculation of flow field with the CFD solver, such as the large eddy simulation (LES), direct numerical simulation (DNS) tasks, etc., which is a memory demanding, computationally expensive and time-consuming iterative process \cite{guo2016convolutional}.

In recent years, with the development of computer technology and the improvement of hardware resources, machine learning and deep learning technology have achieved great success in computer vision \cite{LocalNet, DBLP, ibtehaz2020multiresunet, liang2021fast}, natural language processing (NLP) \cite{wang2021covid, bragg2021flex, lewis2020retrieval, tetko2020state}, speech recognition translation \cite{wang2020transformer,han2020contextnet, ravanelli2020multi, zhang2020transformer} and other scenarios. 
Due to the powerful learning capability of the neural network, machine learning technology as the fourth paradigm of studying aerodynamics, which has also attracted widespread attention in recent years.
Compared with the CFD, machine learning methods only needs to spend a certain amount of time in the early stage to train the neural network, then use the trained neural network model in a few seconds or even milliseconds to get the prediction results of the airfoil flow field.
Guo et al. \cite{guo2016convolutional} used convolutional neural networks (CNN) for variable geometry flow field prediction. 
The test results show that CNN can effectively predict the whole velocity field of geometry. 
When calculating the velocity field, CNN are four orders of magnitude faster than CPU-based solvers and two orders of magnitude faster than GPU-accelerated solvers. 
Thuerey et al. \cite{thuerey2020deep} used U-Net deep learning models instead of Reynolds-Averaged Navier-Stokes (RANS) solvers to solve for pressure and velocity distributions around different airfoils. 
The U-Net architecture is similar to a special codec, under the NVIDIA GTX 1080 GPU platform, U-Net takes about $5.53ms$ to calculate the flow field of airfoil, but OpenFOAM solver takes $40.4s$ to compute the same airfoil. 
Chen et al. \cite{chen2020flowgan} used a generator based on U-Net architecture to generate predictions results of flow fields. 
The multi-layer perceptron (MLP) is used to merge geometry information and flow parameters. 
Combining conditional generative adversarial network (cGAN) and U-Net can establish the mapping relationship between geometry shape and flow fields.
The method obtains good prediction results on the large-scale test set. 

Most CNN-based flow field prediction methods use a data pre-processing method that projects flow field data into a uniformly distributed Cartesian grid \cite{peng2020time, hui2020fast, li2021novel, hu2022mesh}. 
This treatment is feasible for the flow field of interest not to contain geometric features. 
However, when the flow area contains geometric features, pixelation will inevitably cause the lack of geometric information, making it difficult to characterize the flow field details of the near wall area, and even generating non-physical solutions inside the geometry. 
To describe the geometric characteristics of the airfoil, Sekar et al. \cite{sekar2019fast} used deep learning techniques to replace the traditional airfoil parameterization process, using seventy parameters to characterize the airfoil, and the method has good generalization even as the number of airfoil samples increases. 
Because the process of airfoil parameterization is independent of flow field prediction, therefore, it can debugged separately as a module. 
In this study, we use a similar neural network architecture to characterize the airfoils in the UIUC airfoil coordinates database \cite{uiuc}.
The airfoil parameterization network is accomplished by PyTorch neural network framework \cite{PyTorch}. 

For the problem of flow field prediction, a large number of training samples are needed to obtain more accurate prediction results.
For example, in order to predict the flow field around 110 NACA airfoils, Sekar et al. \cite{sekar2019fast} randomly obtained 5280 flow field cases under different Reynolds numbers and angles of attack for neural network model training. 
If each case contains 12000 data points, the training data reaches a staggering $60 \times 20 \times 10^6$ data points, although more training samples will improve the generalization, but the huge training data increases the training time of neural network model. 
According to the experimental results of Sekar et al., under the CPU architecture, the MLP training time up to 1440 hours. 
Nagawkar et al. \cite{nagawkar2022multifidelity} use random forest (RF)-based algorithm to predict high-fidelity flow field, the results show that RF can well predict the pressure and skin friction coefficients of RAE2822 airfoil.

Most of the previous research work was carried out with sufficient training samples. 
Nevertheless, due to the high expenditure of CFD simulations and wind tunnel tests, it is often impossible to obtain sufficient training samples covering all flow conditions \cite{haizhou2022generative}.
Such that the trained neural network cannot guarantee prediction accuracy when the amount of training data is insufficient.
In this work, we propose a multi-head perceptron (MHP) neural network to predict the incompressible steady flow field for the airfoil with sparse samples.
For small sample set with sparse data, if perform multi-variable prediction tasks, the traditional MLP method needs to balance multiple variables during the parameter update process of neural network back-propagation, which leads to model distraction.
The MHP with multiple sub-networks is used to predict aerodynamic parameters with different distribution characteristics. So that the proposed model will pay more attention to the sparse flow field parameters during the back propagation of neural network. 
Secondly, by decoupling the prediction tasks of different parameters, the interference in the prediction process of different parameters is avoided.
Experiments are conducted to evaluate the airfoil flow field prediction accuracy and training time of MLP and MHP.

	%
	%
	%
The rest of the paper is organized as follows. 
Section \uppercase\expandafter{\romannumeral2} mainly describes the airfoil flow field prediction problem and deep learning methods of flow field prediction. 
Section \uppercase\expandafter{\romannumeral3} discusses the data pre-processing problem in airfoil parameterization and flow field prediction. 
Section \uppercase\expandafter{\romannumeral4} shows and discusses the results of neural network model training and prediction. 
And the conclusion is given in Section \uppercase\expandafter{\romannumeral5}.

\section{Methodology}

\subsection{Problem description} 

\begin{figure*}[!h]
	\begin{center}
		\includegraphics[width=0.8 \linewidth]{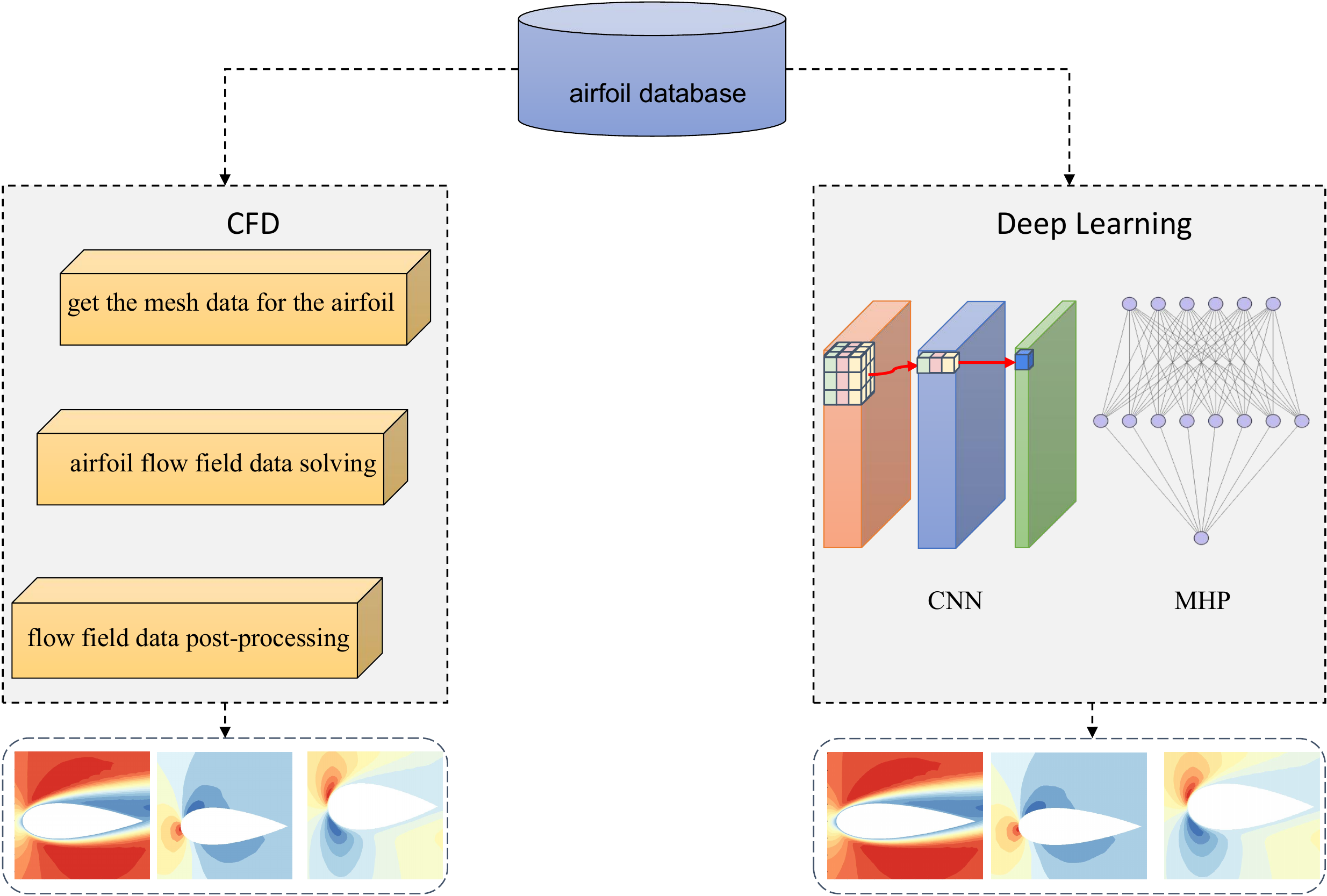}
	\end{center}  \vspace{-2mm}  
	\caption{{{Computational process of CFD simulation and deep learning method}
	}} \label{overview} 
\end{figure*}

The traditional CFD methods to calculate the flow field needs to be meshed according to the initial airfoil coordinates. 
The CFD solver such as Fluent, OpenFOAM, etc. is used to calculate the flow field information around the airfoils, and the calculation results can be displayed by post-processing software such as Tecplot, etc.
The above process only involves a single airfoil, and it may take one hour or more to obtain the final flow field. 
Because CNN and MHP have powerful feature extraction capability and nonlinear fitting capability, in this study, they are used to make geometric parameterization and flow field prediction of airfoil, respectively.
Compared with the traditional CFD calculation method, the pre-trained MHP model can obtain the flow field prediction results of the airfoil in a few seconds. More comparison details can be found in Fig. \ref{overview}.

\subsection{Deep learning methods}

\subsubsection{Convolutional neural network} 

\begin{figure*}[!h]
	\begin{center}
		\includegraphics[width=0.9 \linewidth]{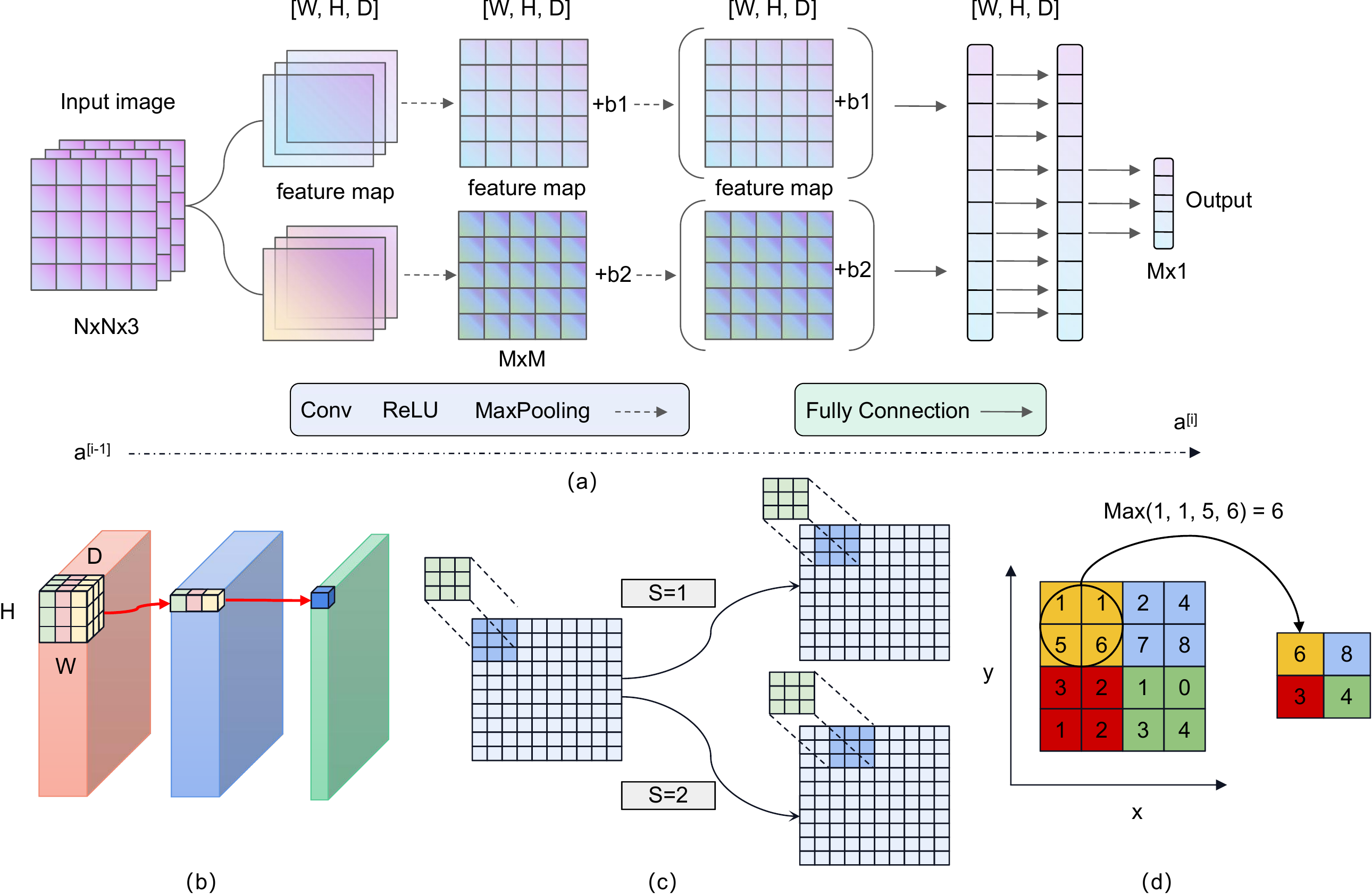}
	\end{center}  \vspace{-2mm}  
	\caption{{{Typical convolutional neural network}
	}} \label{convolution_net} 
\end{figure*}

As the basic component of feature extraction, convolutional neural network have been widely used in well-known neural network architectures such as VGG16 \cite{simonyan2014very}, ResNet \cite{he2016deep}, Faster-RCNN \cite{ren2015faster}, YOLO \cite{redmon2016you}, and SSD \cite{liu2016ssd}. 
As shown in Fig. \ref{convolution_net}(a), a typical convolutional neural network consists mainly of an input layer, a convolutional layer, an activation function layer, a pooling layer, a fully connected layer and an output layer. 
In the Fig. \ref{convolution_net}(a), $W$, $H$ and $D$ represent the width, height and depth of the image, respectively.
$b$ represents the bias of the layer $i$ convolution kernel, in which the dotted line indicates that the image has made corresponding calculation operations through the convolution layer, activation function and max pooling layer. 
The solid line indicates that the data has made relevant calculation operations through the fully connected layer. 
The calculation process of the fully connected layer here is the same as that of the MLP. 
For more details about the MLP, please refer to Section 2.2.3. 
Figure \ref{convolution_net}(b) shows the convolution operation of the convolution kernel. 
The convolution kernel calculates the image on multiple channels through the calculation method of the sliding window to obtain the image feature map. 
The common convolution kernel size are 3 $\times$ 3 and 5 $\times$ 5. 
Figure \ref{convolution_net}(c) shows the influence of step size selection on the calculation results during convolution operation. 
The primary function of the activation function is to provide the nonlinear modeling capability of the network. 
In this work, the ReLU activation function is used in the after convolutional layer and the Tanh activation function is used in the after fully connected layer. 
Max pooling layer is used to extract the principal features of a certain region, reduce the number of parameters and prevent the model from over fitting. 
More details can be found in Fig. \ref{convolution_net}(d). Because the size of the image will decrease after convolution, in order to convolute the image for many times, can fill a specific value around the input matrix, which is generally zero by default. 
Therefore, padding values also affects the size of the output matrix, its width and height can be calculated by the following formula, where $W$ and $H$ represent the width and height of the input image, respectively. 
The $P_i$ is the padding size.
The $K_i$ is the size of the corresponding convolutional kernel, and the $S_i$ is the stride of the convolution kernel:

\begin{equation}
\left\{
\begin{aligned}
H_{out} &= \frac{H + 2 \times P_0 - K_0}{S_0} + 1, \\
W_{out} &= \frac{W + 2 \times P_1 - K_1}{S_1} + 1.
\end{aligned}
\right.
\end{equation}

\subsubsection{Airfoil parameterization network}

\begin{figure*}[!h]
	\begin{center}
		\includegraphics[width=0.9 \linewidth]{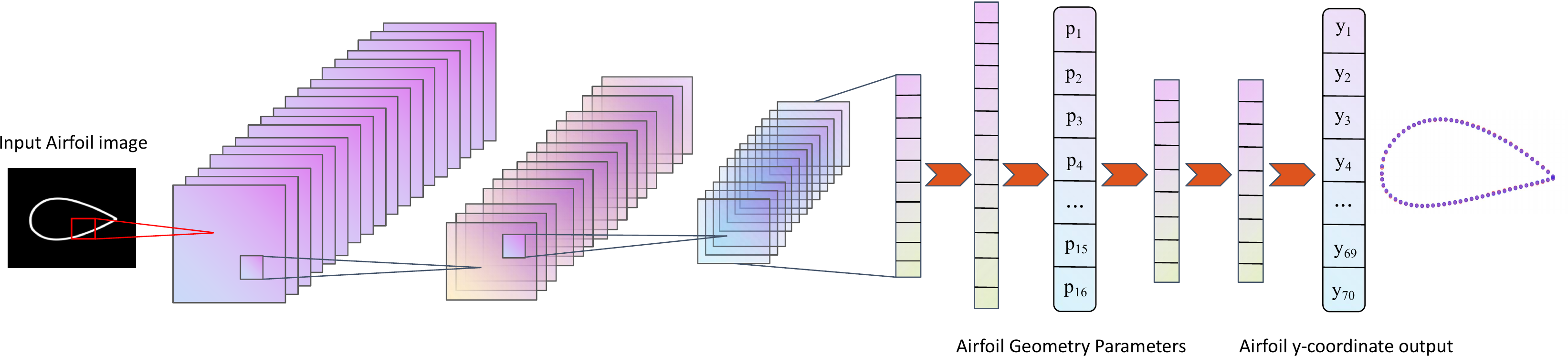}
	\end{center}  \vspace{-2mm}  
	\caption{{{Geometric parameterization network of airfoil}
	}} \label{convolution_parametric} 
\end{figure*}

Because CNN has the property of weight sharing, they have a greater advantage over MLP when performing airfoil image calculations. 
In Fig. \ref{convolution_parametric}, the airfoil parameterization network is composed of CNN and MLP. 
They are used to encode the input airfoil image into sixteen important geometric parameters and then decode these parameters into the $y$ coordinates corresponding to the current airfoil image. 

Refer to the experimental results of Sekar et al. \cite{sekar2019fast}, and the influence of the number of convolutional layers and fully connected layers on the model training results is not considered. 
The details of the network architecture are shown in Tbl. \ref{cnn_architecture}. 
$Cov_{i}$ represents the layer $i$ convolutional, and $Fcn_{j}$ represents the layer $j$ fully connected.
The first product in the fourth column represents the convolutional kernel size, and the second product represents the max pooling layer filter size. 
Each convolutional layer is followed by a ReLU activation function, and a Tanh activation function followed the max pooling layer. 
Using the mean squared error (MSE) as the loss function of the model, we define it as:

%
\begin{equation}
CNN_{loss} = \frac{1}{N} \sum_{i=1}^{N} \sum_{j=1}^{70} (y_{i,j}^t - y_{i,j}^p)^2
\end{equation}
\begin{table}[tb]
	\caption{Airfoil parametrization convolutional neural network architecture} \label{cnn_architecture}
	\begin{center}
		\begin{tabular}{lccc}
			\hline \hline
			Layer type & In channels & Out channels & Kernel size \\ \hline
			Cov1      & 1            & 32             & 4$\times$4, 3$\times$3            \\
			Cov2       & 32            & 32             & 4$\times$4, 3$\times$3            \\
			Cov3       & 32            & 64             & 4 $\times$4, 3$\times$3          \\
			Cov4       & 64          &  64             & 4 $\times$4, 2$\times$2           \\
			Cov5       & 64             &  128             & 4 $\times$4, 2$\times$2           \\
			Fcn1        &  128            &   100           &              \\
			3$\times$Fcn2        & 100             & 100        &              \\
			
			Fcn3     & 100             &  16             &              \\ 
			Fcn4     & 16          & 100       &        \\
			3$\times$Fcn5 & 100  &100  &  \\
			Output &100 & 70 &  \\
			\hline \hline
		\end{tabular}
	\end{center} 
\vspace{-1.5em}
\end{table} 
Equation (2) represents the loss function of a batch, which is set to 64.
$y_{i,j}^{t}$, $y_{i,j}^{p}$ represents the ground-truth and predicted values of the $y$ coordinate of the airfoil, respectively.

\subsubsection{Multi-layer perceptron}
A typical multi-layer neural network is shown in Fig. \ref{mlp}(a), which consists mainly of three parts: input layer, hidden layer and output layer. 
The input of the MLP is twenty physical parameters. 
$G_i$ represents the geometric parameters calculated by CNN parameterization network, which are used to characterize the geometric shape of different airfoils. 
The Reynolds number along with the angle of attack is used to describe the physics field information in which the current airfoil is located. 
The $x$ coordinate and $y$ coordinate are used to illustrate coordinate information for different points in the airfoil flow field. 
The MLP neural network is fully connected between the different layers, that is,  it connected any neurons in the upper layer to all neurons in the next layer. 
MLP have three basic elements: weights, biases, and activation functions. 
Weights control the strength of the connections between neurons, the size of which indicates the magnitude of the likelihood. 
The bias is set to correctly classify the sample and is an important parameter in the model, which is to ensure that the output values calculated from the input values cannot be activated casually. 
Activation functions act as nonlinear mappings that limit the output amplitude of neurons to a certain range, generally between $(-1, 1)$ or $(0, 1)$. 
The output of the neural network is pressure and velocity in the $x$ and $y$ directions, respectively.
The prediction function of the MLP can be defined as:
\begin{equation}
f_{MLP}(p_{1},...,p_{16},Re,AOA,x,y) = (u,p,v) .
\end{equation}
The left side of the function represents the prediction model of the MLP, and the right side of the function represents the prediction result of the model.

MSE is used as the loss function of MLP, which is defined as:
\begin{equation}
MLP_{loss} = \frac{1}{3 \times N} \sum_{i=1}^{N}[(u_{i}^{t} - u_{i}^{p})^2 + (p_{i}^{t} - p_{i}^{p})^2 +(v_{i}^{t} - v_{i}^{p})^2] ,
\end{equation}
where $u_{i}^{t}$ represents the ground-truth of the velocity component in the $x$ direction, and $u_{i}^{p}$ represents the predicted values of the velocity component in the $x$ direction.
$v_{i}^{t}$ represents the ground-truth of the velocity component in the $y$ direction, and $v_{i}^{p}$ represents the predicted values of the velocity component in the $y$ direction.
$p_{i}^{t}$, $p_{i}^{p}$ represents the ground-truth and prediction value of pressure, respectively. 

\subsubsection{Multi-head perceptron}

\begin{figure*}[!h]
	\begin{center}
		\includegraphics[width=1 \linewidth]{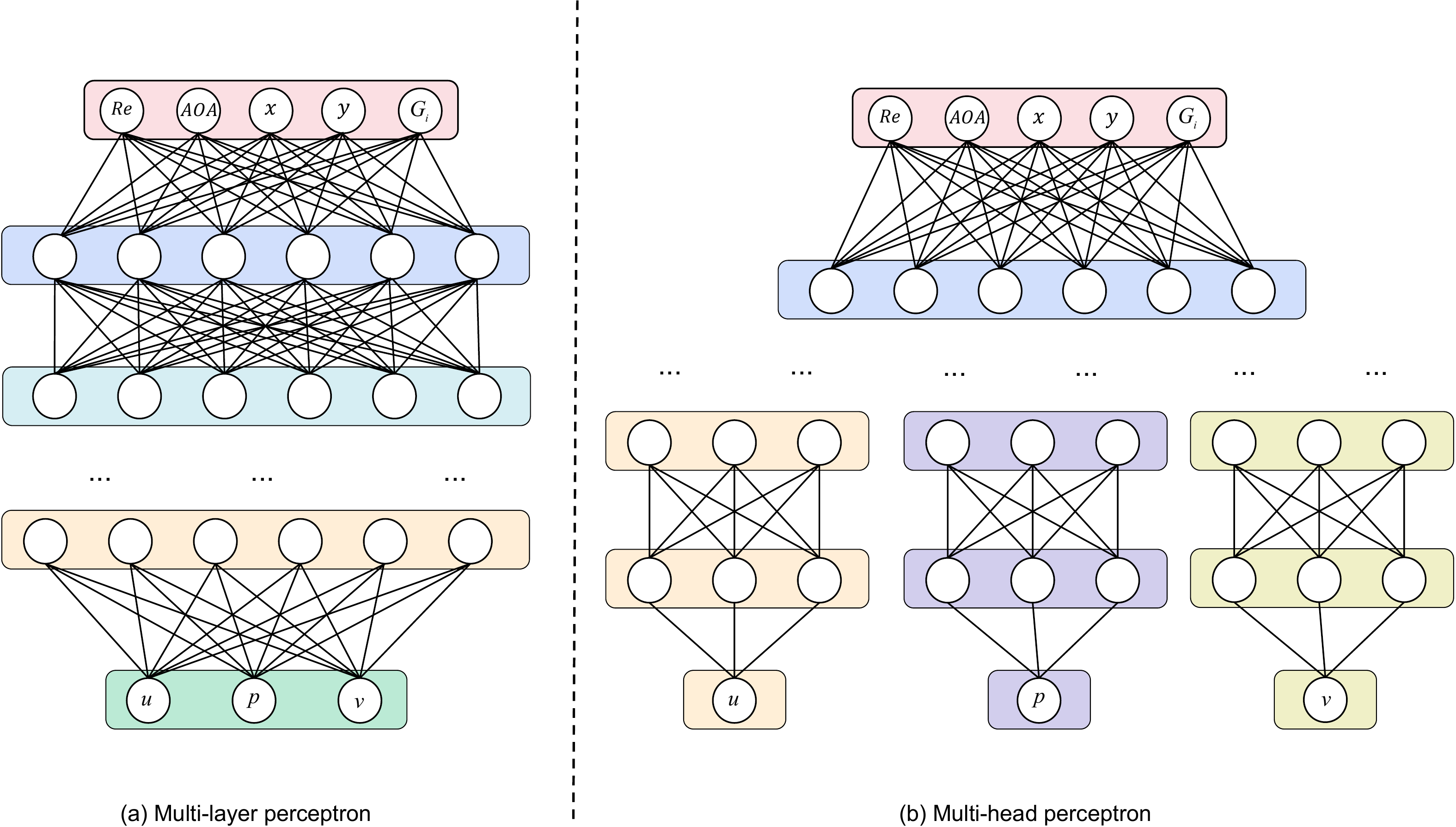}
	\end{center}  \vspace{-2mm}  
	\caption{{{Neural networks for flow field prediction}
	}} \label{mlp} 
\end{figure*}

Most of the previous research works \cite{sekar2019fast, yu2019flowfield, pawar2021physics} has adopted the network architecture shown in Fig. \ref{mlp}(a).
But it can be found through related experimental results (more details can be found in Section IV), MLP has obvious drawbacks in processing sparse data.
Because for sparse data with uneven distribution, even if the training data is normalized, it is also difficult for MLP to produce enough observations.
Especially in the case of multiple outputs, in order to maximize the prediction of multiple target values, MLP will update the network parameters of each previous layer in back-propagation.
Due to the existence of sparse data, the final fitting effect of MLP is unsatisfactory. 
In Fig. \ref{mlp}(b), in order to avoid the interference of sparse data on other aerodynamic parameters to be predicted, MHP is proposed to predict various physical parameters in the flow field, respectively. 
In particular, firstly, the basic network is used to extract the flow field characteristics, and then the multi-head network is used to obtain the prediction output of the network for three physical parameters with different distribution characteristics.
The prediction function of the MHP can be defined as:
\begin{equation}
\left\{
\begin{aligned}
f_{MHP}(p_{1},...,p_{16},Re,AOA,x,y) &= (u) , \\
f_{MHP}(p_{1},...,p_{16},Re,AOA,x,y) &= (p) , \\
f_{MHP}(p_{1},...,p_{16},Re,AOA,x,y) &= (v) .
\end{aligned}
\right. 
\end{equation}

The left side of the function represents the prediction model of the MHP, and the right side of the function represents the prediction result of the model.
MSE is also used as the loss function of the MHP. The loss function of each head of MHP is defined as:

\begin{equation}
\left\{
\begin{aligned}
MHP(u)_{loss}=\frac{1}{N}\sum_{i=1}^{N}(u_{i}^{t}-u_{i}^{p}), \\
MHP(p)_{loss}=\frac{1}{N}\sum_{i=1}^{N}(p_{i}^{t}-p_{i}^{p}), \\
MHP(v)_{loss}=\frac{1}{N}\sum_{i=1}^{N}(v_{i}^{t}-v_{i}^{p}). 
\end{aligned}
\right. 	
\end{equation}

\section{Data preparation}

\subsection{Airfoil dataset}

\begin{figure*}[!h]
	\begin{center}
		\includegraphics[width=0.6 \linewidth]{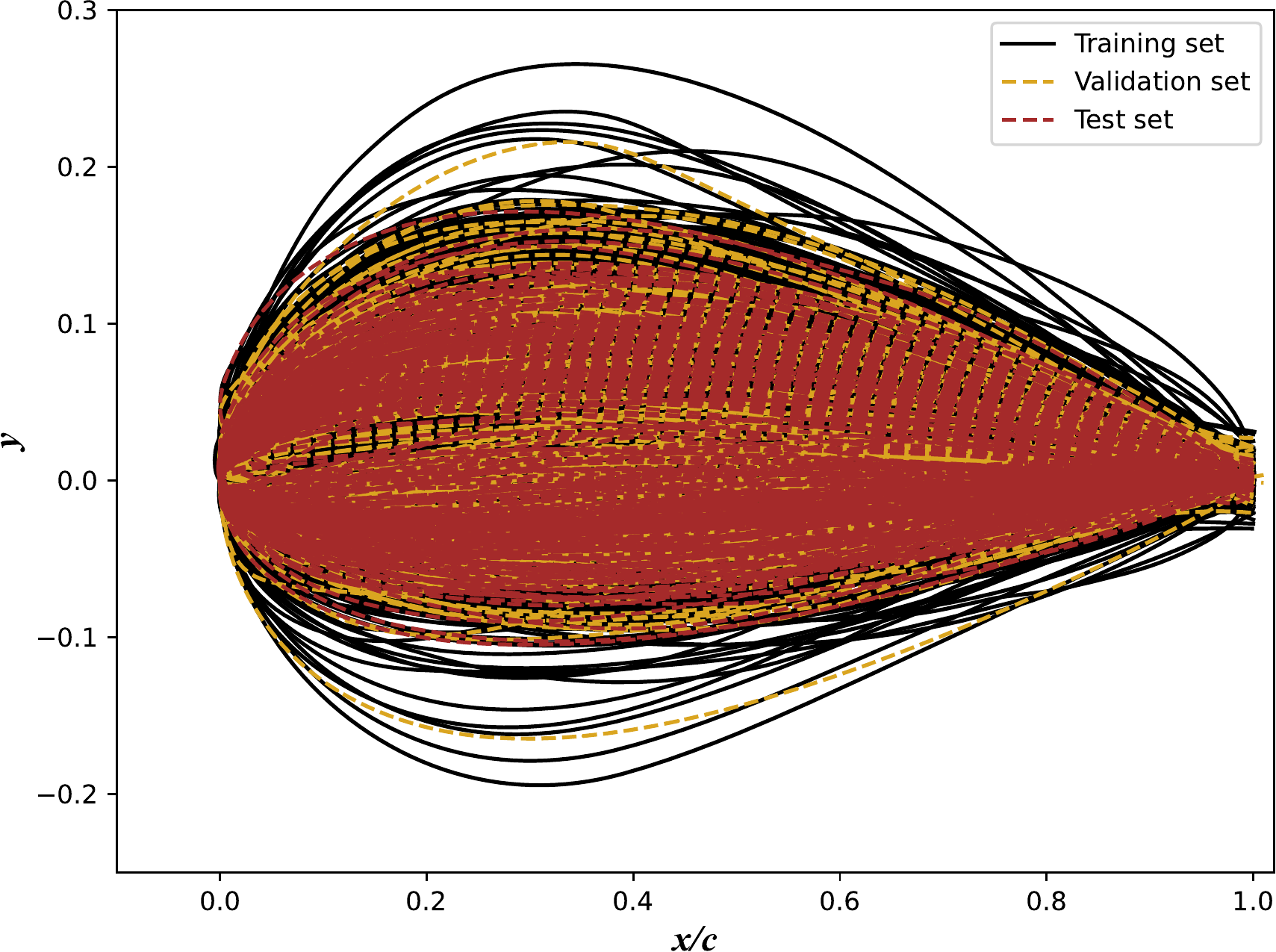}
	\end{center}  \vspace{-2mm}  
	\caption{{{Airfoil shapes used in airfoil parameterization network.}
	}} \label{uiuc_train} 
\end{figure*}

\begin{figure*}[!h]
	\begin{center}
		\includegraphics[width=0.8 \linewidth]{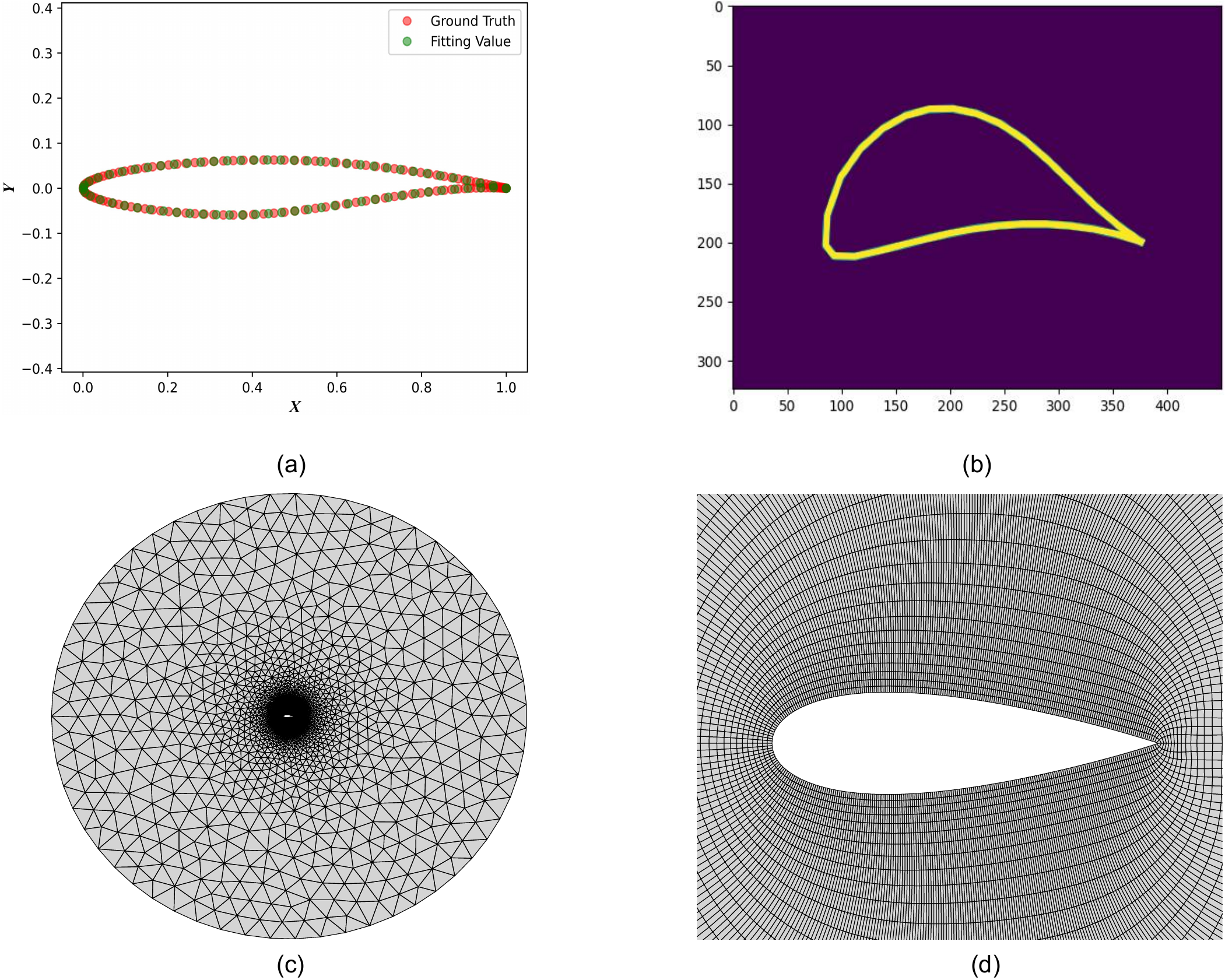}
	\end{center}  \vspace{-2mm}  
	\caption{{{Schematic diagram of data pre-processing. $(a)$ Fitting curve of RAE2822. $(b)$ Airfoil image for CNN training. $(c)$ Full flow field mixed mesh for CFD calculation. $(d)$ Structured mesh of airfoil near wall. }
	}} \label{uiuc} 
\end{figure*}

The CNN network architecture is used to predict the airfoil geometric parameters in the UIUC airfoil database. The airfoil shapes for CNN training is shown in Fig. \ref{uiuc_train}.
Because the number of $x$ coordinate and $y$ coordinate in the airfoil UIUC database is not uniform.
In order to facilitate the training of neural network models, each airfoil is first fitted using the nonuniform rational B-spline (NURBS) method \cite{lepine2001optimized,lepine2000wing}, and then seventy data points are randomly selected on the fitting airfoil curve as the new $x$ coordinate and $y$ coordinate of the current airfoil. 
The fitting results of RAE2822 is shown in Fig. \ref{uiuc}(a). 
Secondly, fix the $x$-coordinate along the chord length, each data point of $x$-coordinate is obtained by following calculation formula:

\begin{equation}
x_{j}=\frac{1}{n} \sum_{i=1}^{n} x_i
\end{equation}

Then select the normalized $y$ coordinate as the target label value of the airfoil image at model training, the numeric range of the normalized $y$ coordinate is $(-1,1)$.
The calculation formula is following:

\begin{equation}
y_{j} = \frac{y_i - y_{avg}}{y_{max} - y_{min}}
\end{equation}
 
In the above formula, $y_{avg}$ represents the global average, $y_{max}$ represents the global maximum, $y_{min}$ represents the global minimum. 
The normalized $y$ coordinate and fixed $x$-coordinate can be used to obtain images of different airfoils and each airfoil image also needs to be normalized. 
First of all, the gray scale image with a single channel size of 216 $\times$ 216 needs to be inverted and normalized, so that the pixel value on the airfoil geometry curve is $1$, the pixel value that is not on the airfoil geometry curve is $0$, and the rest of the pixel values are between $0$ and $1$. 
The results of the airfoil image pre-processing is shown in Fig. \ref{uiuc}(b).
Eighty percent of the 1582 airfoils in UIUC airfoil database are used for model training set, $10\%$ for cross validation set, and the remaining $10\%$ are used as the test set.

\subsection{Flow field dataset}
MHP and MLP is used to predict the flow field around the airfoils. 
The airfoil flow field database is generated by CFD method.
To verify the prediction effect of MHP and MLP in the case of sparse samples.
Here for the NACA0006, NACA0008, NACA0012, NACA0024 four airfoils in the range of Reynolds number 1000 to 2000 (Re: 1000, 1200, 1800, 2000), angle of attack 0\degree \, to 10\degree \, (AOA: 0\degree, 2\degree, 4\degree, 6\degree, 8\degree, 10\degree).
Each airfoil generates $24$ cases, a total of $96$ cases as the training set and the test set of MLP and MHP (of which $80\%$ of the data is used as the training set, $10\%$ of the data is used as the cross-validation set, and $10\%$ of the data is used as the test set). 
As shown in Fig. \ref{uiuc}(c) and Fig. \ref{uiuc}(d), only the flow field of the structured mesh around the airfoil is selected as the training data and test data of the model, and the far-field unstructured mesh data does not participate in the training and prediction of the model. 
Since the input of MLP and MHP is $20$ parameters ($p_{1},..,p_{16}, Re, AOA, x, y$), and the flow field data of a single airfoil is about $8358$ rows, a total of $8358 \times 20$ data of a single airfoil participates, and the final full flow field data is about $8358 \times 20 \times 96$. 
Similar to the parametric network model, in order to accelerate the convergence speed of the model during training, Reynolds number, angle of attack, $x$ coordinates, $y$ coordinates, $u-velocity$, $pressure$, $v-velocity$, are also normalized here.
The input values are normalized to the range of 0 to 1.
The sixteen geometric parameters of the airfoil obtained through the parametric network are not normalized in this research, because their numerical ranges are between -1 and 1. 
In the experiments of Section 4.2, the flow field data of this section are used for relevant test work.
More details about training and prediction results of MLP and MHP can be found in Section \uppercase\expandafter{\romannumeral4}.

\section{Results and discussions}

\subsection{Airfoil parameterization}

CNN is used for parameterization of airfoils. 
The model parameters are optimized using the Adam optimizer.
The initial learning rate is $2.5 \times 10^{-4}$, and the epoch is set to 5000, which means that the neural network traverses all the training data 5000 times during training. 
The program is implemented using PyTorch deep learning framework, and the GPU (RTX3060) is used for the model training under the Linux platform. 
From Tbl. \ref{cnn_device}, it can be found that the training speed of the model can be greatly improved by using the GPU.
The training time of the airfoil parametric network under the CPU is about 53.5h, while the training time with the GPU is only 1.5h. 

\begin{figure*}[!h]
	\begin{center}
		\includegraphics[width=0.6 \linewidth]{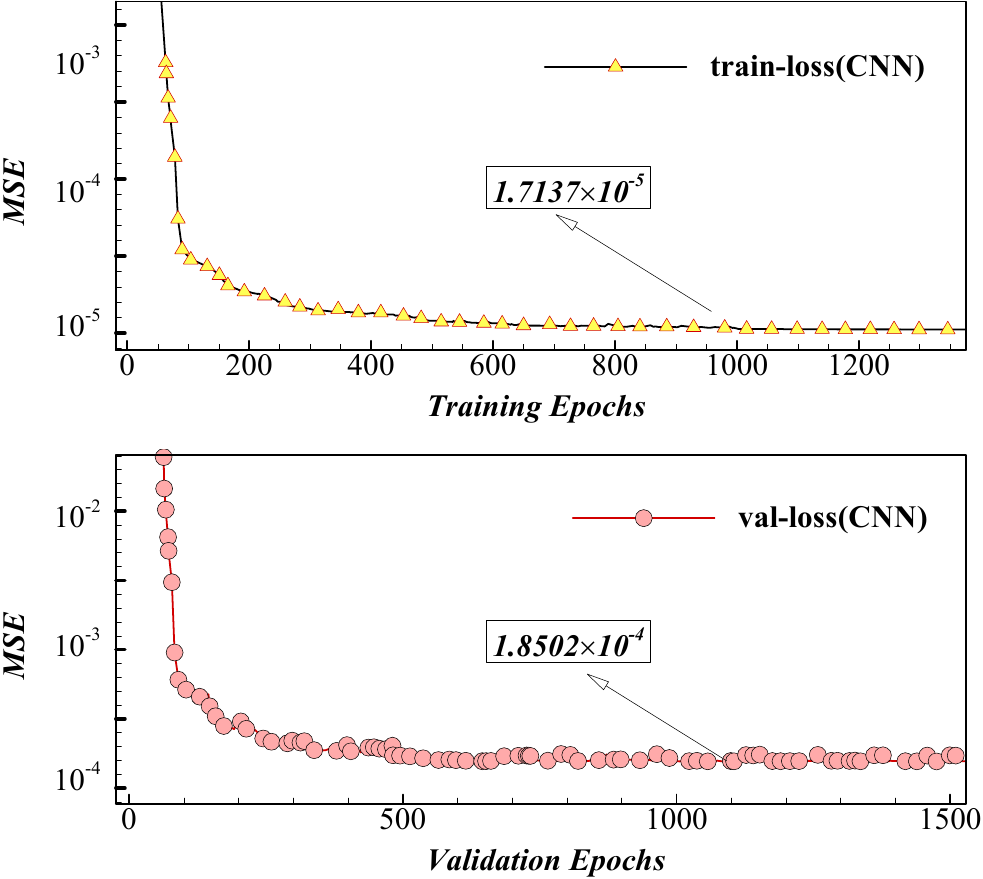}
	\end{center}  \vspace{-2mm}  
	\caption{{{CNN loss function curve}
	}} \label{cnn_loss} 
\end{figure*}

\begin{figure*}[!h]
	\begin{center}
		\includegraphics[width=0.8 \linewidth]{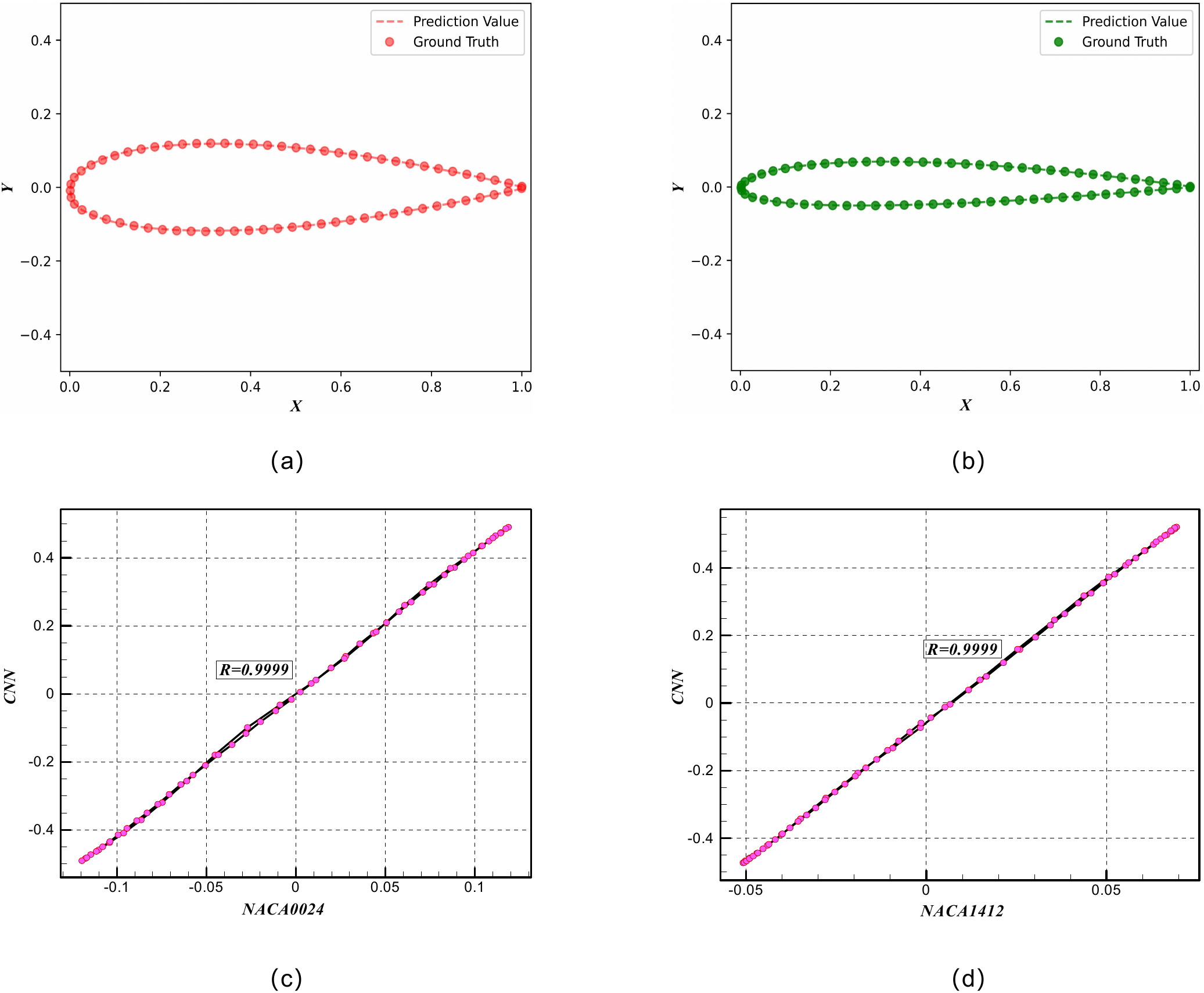}
	\end{center}  \vspace{-2mm}  
	\caption{{{CNN prediction results and correlation curves. (a) CNN prediction result of NACA0024. (b) CNN prediction result of NACA1412. (c) Correlation curve of NACA0024. (d) NACA1412 correlation curve.}
	}} \label{cnn_relative_co} 
\end{figure*}

Figure \ref{cnn_loss} shows the loss function curve of training set and cross-validation set during the CNN training.
After 1000 epochs, the model has basically reached the convergence state.
The loss function on the training set eventually converges to $1.7137\times10^{-5}$, and the loss function on the cross-validation set eventually converges to $1.8502\times10^{-4}$.
Figure \ref{cnn_relative_co}(a) and Fig. \ref{cnn_relative_co}(b) shows the fitting results of prediction values and ground truth of NACA0024 and NACA1412, respectively.
In Fig. \ref{cnn_relative_co}, the predicted value of CNN has a good fitting effect with the ground truth, indicating that the geometric parameters obtained in the airfoil parametric network can well characterize the current airfoil geometry shape.
The prediction accuracy of the CNN is further verified by using the correlation coefficient between ground truth and prediction value, which is defined as:

%
\begin{equation}
	R=\frac{cov(T,P)}{\sigma_{T}\sigma_{P}}=\frac{\sum_{i=1}^{n}(T_{i}-\bar{T})(P_{i}-\bar{P})}{\sqrt{\sum_{i=1}^{n}(T_{i}-\bar{T})^2}\sqrt{\sum_{i=1}^{n}(P_{i}-\bar{P})^2}}
\end{equation}
%

In the above formula, $cov$ represent the covariance and $\sigma$ is the standard deviation.
$T$ and $P$ represent the ground truth of $y$ coordinates and the prediction values of CNN, respectively. 
$\bar{T}$ and $\bar{P}$ represent the average of $T$ and $P$, respectively.
As shown in Fig. \ref{cnn_relative_co}(c), the correlation coefficient R=0.9999 between prediction value and ground truth of the NACA0024. 
In Fig. \ref{cnn_relative_co}(d), the correlation coefficient R=0.9999 between prediction value and ground truth of NACA1412. 
The discrete points of the two images are distributed near the diagonal, indicating that the degree of coincidence between the prediction value and the ground truth is high.
CNN has achieved a good prediction effect on the data in the airfoil database.

%
\begin{table}[tb]
	\caption{Environment configuration of airfoil parameterization network} \label{cnn_device}
	\begin{center}
		\begin{tabular}{lcc}
			\hline \hline
			Name     & version                                                           & train time \\ \hline
			platform & Linux                                                             &            \\
			CPU      & \begin{tabular}[c]{@{}l@{}}Intel i7-11700K 3.6GHz\\  \end{tabular} & 53.5h      \\
			GPU      & RTX3060                                                           & 1.5h       \\
			cuda     & 11.6                                                              &            \\
			PyTorch  & 1.11.0+cu113                                                      &            \\ 
			\hline \hline
		\end{tabular}
	\end{center} 
\vspace{-1.5em}
\end{table} 

\subsection{Flow field prediction}

\subsubsection{MLP training results}
\begin{figure*}[!h]
	\begin{center}
		\includegraphics[width=1 \linewidth]{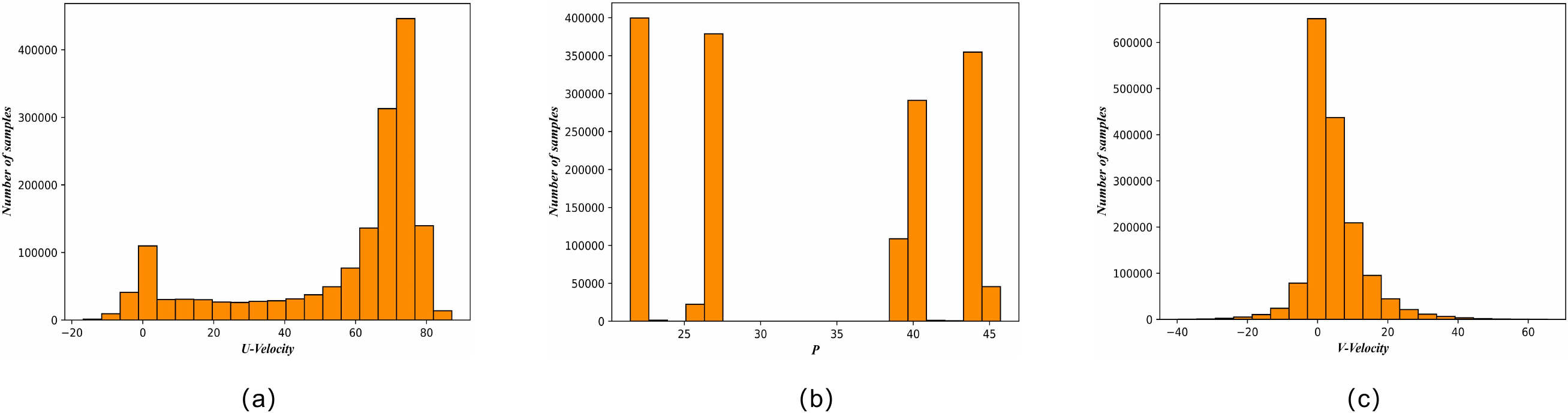}
	\end{center}  \vspace{-2mm}  
	\caption{Histogram of training set. (a) Distribution histogram of U-velocity. (b) Distribution histogram of pressure. (c) Distribution histogram of V-velocity.} \label{hist} 
\end{figure*}

\begin{figure*}[!h]
	\begin{center}
		\includegraphics[width=1 \linewidth]{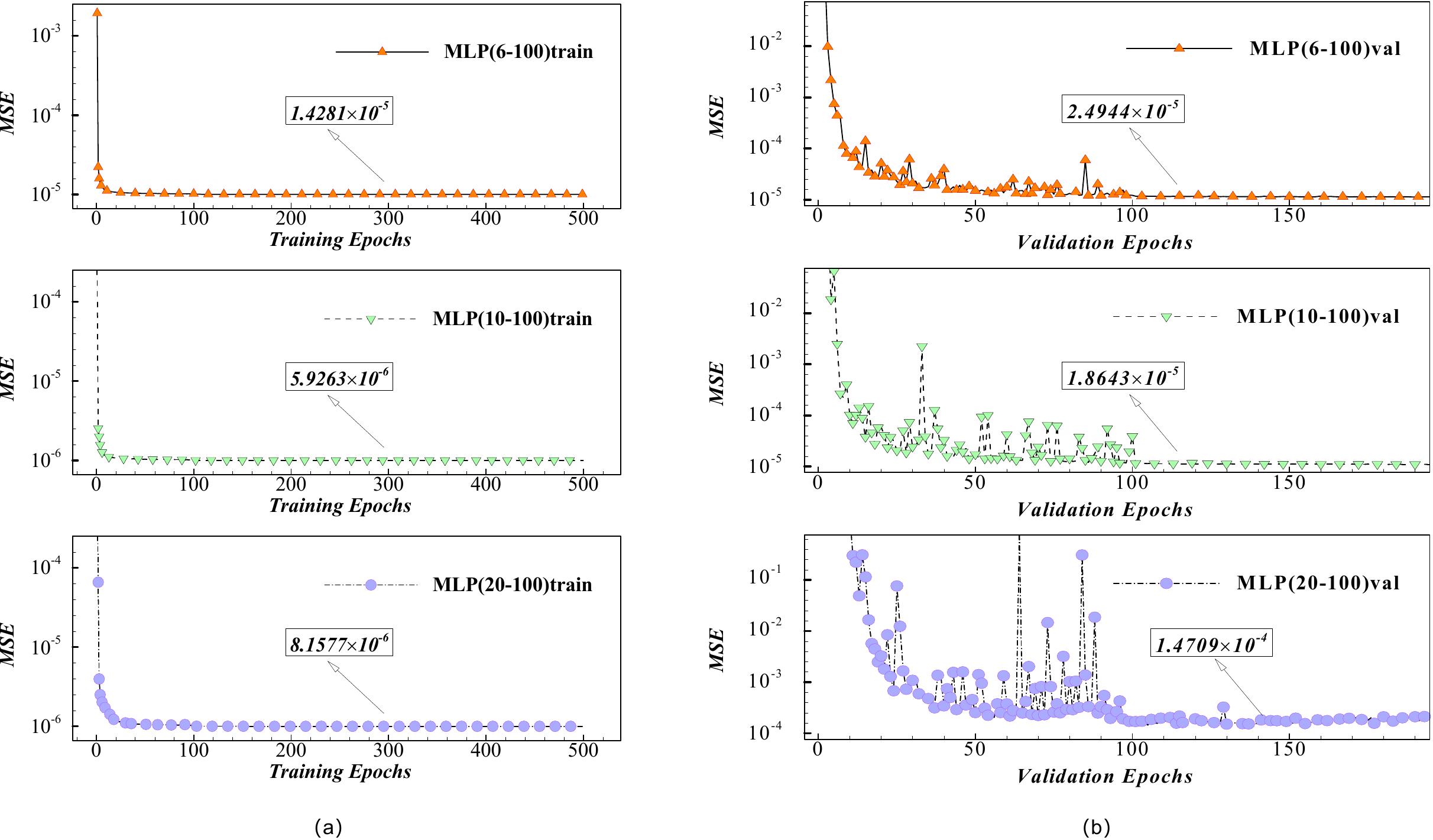}
	\end{center}  \vspace{-2mm}  
	\caption{{{MSE convergence curves of MLP with different layers. (a) MSE convergence curve of MLP with different layers on training set. (b) MSE convergence curve of MLP with different layers on cross-validation set}
	}} \label{multilayer mlp} 
\end{figure*}

\begin{figure*}[!h]
	\begin{center}
		\includegraphics[width=1 \linewidth]{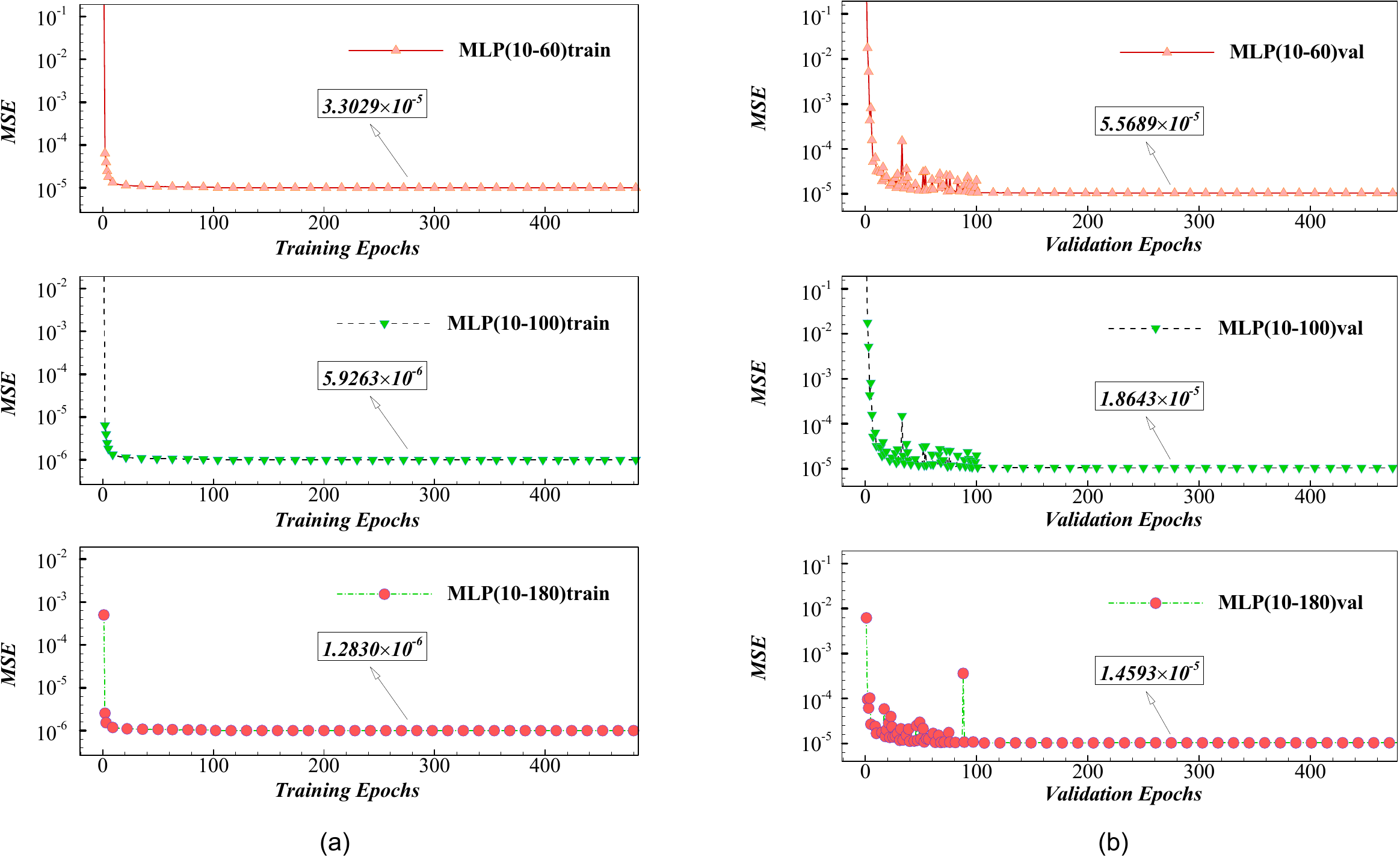}
	\end{center}  \vspace{-2mm}  
	\caption{{{MSE convergence curves of MLP with different nodes. (a) MSE convergence curve of MLP with different nodes on training set. (b) MSE convergence curve of MLP with different nodes on cross-validation set}
	}} \label{multilnode mlp} 
\end{figure*}

\begin{table}[tb]
	\caption{Test network architectures with same nodes and different layers} \label{multi_layer mlp}
	\begin{center}
		\begin{tabular}{lccc}  
			\hline \hline
			Name & MLP(6-100) & MLP(10-100) & MLP(20-100)  \\ \hline
			Input      & 1 $\times$ 20          & 1 $\times$ 20           & 1 $\times$ 20             \\
			Hidden     & 6 $\times$ 100         & 10 $\times$ 100           & 20 $\times$ 100             \\
			Output     & 1 $\times$ 3          & 1 $\times$ 3           & 1 $\times$ 3           \\ 
			\hline \hline
		\end{tabular}
	\end{center} 
\vspace{-1.5em}
\end{table} 
\begin{table}[tb]
	\caption{Test network architectures with different nodes and same layers} \label{multi_node mlp}
	\begin{center}
		\begin{tabular}{lccc}  
			\hline \hline
			Name & MLP(10-60) & MLP(10-100) & MLP(10-180)  \\ \hline
			Input      & 1 $\times$ 20          & 1 $\times$ 20           & 1 $\times$ 20            \\
			Hidden     & 10 $\times$ 60         & 10 $\times$ 100           & 10 $\times$ 180             \\
			Output     & 1 $\times$ 3          & 1 $\times$ 3           & 1 $\times$ 3            \\ 
			\hline \hline
		\end{tabular}
	\end{center} 
\end{table} 
\begin{table}[tb]
	\caption{Training time and loss function accuracy of MLP with different layers} \label{multi_layer_time}
	\begin{center}
		\begin{tabular}{c|clc}
			\hline \hline
			\multirow{2}{*}{Name} & \multicolumn{2}{c}{MSE}                                        & \multicolumn{1}{l}{Time} \\ \cline{2-4} 
			& training                      & \multicolumn{1}{c}{validation} & \multicolumn{1}{l}{}     \\ \hline
			MLP(6-100)            & $1.4281 \times10^{-5}$                     & $2.4944 \times10^{-5}$                      & 3h8m                     \\
			MLP(10-100)           & \multicolumn{1}{l}{$5.9263 \times10^{-6}$} & $1.8643 \times10^{-5}$                      & 4h26m                    \\
			MLP(20-100)           & \multicolumn{1}{l}{$8.1577 \times10^{-6}$} & $1.4709 \times10^{-4}$                      & 7h59m                    \\ 
			\hline \hline
		\end{tabular}
	\end{center} 
	\vspace{-1.5em}
\end{table} 
\begin{table}[tb]
	\caption{Training time and loss function accuracy of MLP with different nodes} \label{multi_node_time}
	\begin{center}
		\begin{tabular}{c|clc}
			\hline \hline
			\multirow{2}{*}{Name} & \multicolumn{2}{c}{MSE}                                        & \multicolumn{1}{l}{Time} \\ \cline{2-4} 
			& training                      & \multicolumn{1}{c}{validation} & \multicolumn{1}{l}{}     \\ \hline
			MLP(10-60)            & $3.3029 \times10^{-5}$                     & $5.5689 \times10^{-5}$                      & 4h35m                     \\
			MLP(10-100)           & \multicolumn{1}{l}{$5.9263 \times10^{-6}$} & $1.8643 \times10^{-5}$                      & 4h26m                    \\
			MLP(10-180)           & \multicolumn{1}{l}{$1.2830 \times10^{-6}$} & $1.4593 \times10^{-5}$                      & 4h29m                    \\ 
			\hline \hline
		\end{tabular}
	\end{center} 
	\vspace{-1.5em}
\end{table} 
%
Firstly, the traditional MLP method is used to train the flow field of different geometry airfoils under different working conditions. 
Figure \ref{hist} shows the histogram of all training data, in which the velocity value distribution is relatively uniform and the pressure value distribution is more sparse.
The weights of the MLP are trained using the Adam optimizer, the initial learning rate is set to 5 $\times$ $10^{-5}$. 
To accelerate the convergence of the model, the learning rate is optimized using the learning rate scheduler. 
The parameter gamma is set to 0.1, that is, the learning rate is multiplied by 0.1 for every 100 epochs passed. 
The effects of different layers and different nodes on the model training results are tested.
The details of the test network are shown in Tbl. \ref{multi_layer mlp} and Tbl. \ref{multi_node mlp}. 
The training set loss function curve and the validation set loss function curve of different models are shown in the Fig. \ref{multilayer mlp} and Fig. \ref{multilnode mlp}. 
 
In Fig. \ref{multilayer mlp}, the effect of different layers on the loss function accuracy of the MLP is tested. 
From Fig. \ref{multilayer mlp}(a) and Fig. \ref{multilayer mlp}(b), it can be found that the number of MLP neural network layers is increased from 6 layers to 10 layers.
The MSE curve can converge quickly, but the 10-layer MLP can get a smaller MSE and better prediction accuracy.
However, if the number of layers of MLP neural network continues to increase, the convergence rate of the MSE loss function curve is slower.
The MSE of 20-layer MLP is larger than that of 6-layer and 10-layer neural network.
And as can be seen from Tbl. \ref{multi_layer_time}, as the number of layers increases, the training time of the neural network will increase exponentially, but the loss function does not decrease.
Therefore, the 10-layer MLP are selected as the training architecture for subsequent models.
On the training set, the MSE of 10-layer MLP finally converges to $5.9263 \times 10^{-6}$.
On the cross-validation set, the MSE of 10-layer MLP finally converges to $1.8643 \times 10^{-5}$.

As shown in Fig. \ref{multilnode mlp}, the effect of different node numbers on the loss function accuracy of the MLP is tested. 
In Fig. \ref{multilnode mlp}(a) and Fig. \ref{multilnode mlp}(b), the number of nodes in each layer of MLP is increased from 60 to 180.
In the initial training stage of MLP, the MSE decreases rapidly and becomes smaller.
After 100 epochs, MSE basically reached a stable state.
And as can be seen from Tbl. \ref{multi_node_time}, as the number of network nodes increases, the training time does not change much, but the loss function becomes smaller. 
Therefore, in this research, MLP with 10 layers and 180 nodes in each layer is selected as the final flow field prediction neural network architecture. 
In Fig. \ref{multilnode mlp}(a) and Fig. \ref{multilnode mlp}(b), the loss function of the MLP on the training set eventually converges to $1.2830 \times 10^{-6}$, while the loss function on the cross-validation set eventually converges to $1.4593 \times 10^{-5}$.

%
\begin{table}[tb]
	\caption{Training time and loss function of MLP and MHP} \label{MLP and MHP}
	\begin{center}
		\begin{tabular}{cc|ccc}
			\hline \hline
			\multicolumn{2}{c|}{\multirow{2}{*}{Name}}        & \multicolumn{2}{c}{MSE} & Time  \\ \cline{3-5} 
			\multicolumn{2}{c|}{}                             & training  & validation  &       \\ \hline
			\multicolumn{1}{c|}{\multirow{3}{*}{MLP}}     & u & $1.2830 \times 10^{-6}$    & $1.4593 \times 10^{-5}$      & 4h29m \\
			\multicolumn{1}{c|}{}                         & p & $1.2830 \times 10^{-6}$    & $1.4593 \times 10^{-5}$     & 4h29m \\
			\multicolumn{1}{c|}{}                         & v &  $1.2830 \times 10^{-6}$    & $1.4593 \times 10^{-5}$      & 4h29m \\ \cline{1-2}
			\multicolumn{1}{l|}{\multirow{3}{*}{MHP}} & u & $2.4273 \times 10^{-6}$    & $2.6691 \times 10^{-5}$      & 3h42m \\
			\multicolumn{1}{l|}{}                         & p & $4.4581 \times 10^{-8}$    & $1.4126 \times 10^{-7}$      & 4h9m \\
			\multicolumn{1}{l|}{}                         & v & $5.3882 \times 10^{-7}$    & $3.8955 \times 10^{-6}$      & 4h10m \\
			\hline \hline
		\end{tabular}
	\end{center} 
	\vspace{-1.5em}
\end{table} 
\begin{figure*}[!h]
	\begin{center}
		\includegraphics[width=1 \linewidth]{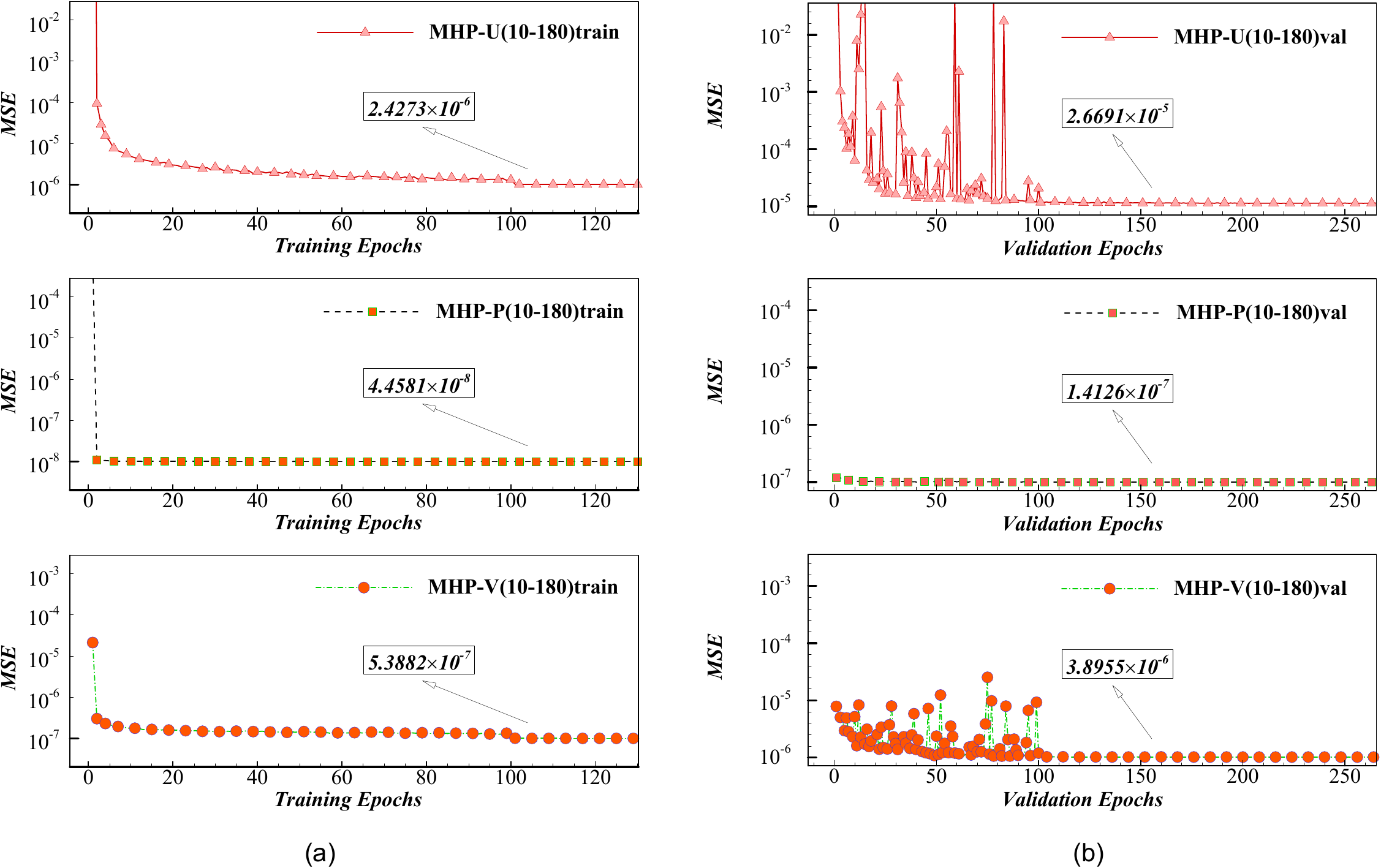}
	\end{center}  \vspace{-2mm}  
	\caption{{{MHP neural network loss function curve. (a) Loss function curve of MHP on training set. (b) Loss function curve of MHP on cross-validation set. }
	}} \label{MHP-net} 
\end{figure*}

\subsubsection{MHP training results}

Secondly, MHP is used to train the flow field data in Section \uppercase\expandafter{\romannumeral3}.
The hyperparameters of the MHP during training, such as learning rate, batch size, number of neural network layers and number of neurons are the same as the MLP. 
And it performed 500 iterations during the model training process. 
Figure \ref{MHP-net} shows the loss function curve of MHP model during training.
As shown in Fig. \ref{MHP-net}(a), on the training set of the flow field, the MHP-U, MHP-P and MHP-V loss function curves all decreased rapidly at the beginning.
The MHP loss function curve decreases the fastest and the value after MSE finally converges is the smallest, about $4.4581\times10^{-8}$. 
The comparison results of MLP and MHP training time and MSE are shown in Tbl. \ref{MLP and MHP}.
Compared with the MLP, the MSE of MHP-U did not change significantly, but the training time of MHP-U was reduced by 47 minutes, while the training time of MHP-P was reduced by 20 minutes and the MSE was reduced by 2 orders of magnitude compared with the MLP.
And the training time of the MHP-V was reduced by 19 minutes and the MSE was reduced by 1 order of magnitude compared with the MLP.
The comparison results show that MHP has more powerful prediction capability than MLP in the face of sparse flow field data.
The loss function curve of the MHP on the cross-validation dataset is shown in Fig. \ref{MHP-net}(b). 
It is found that the MSE of the MHP-P achieves a good result at the beginning of the training.
And the reason why MHP-U and MHP-V curves oscillate before 100 epochs is that the initial learning rate is too large.
And then gradually stabilizes after 100 epochs because the learning rate scheduler is used in this work to automatically reduce the learning rate value during the model training.
Finally, the MHP loss function basically converges after 150 epochs. 
The MSE of MHP-P on the cross-validation set eventually converges to $1.4126 \times 10^{-7}$. 
On the cross-validation set, the loss function of MHP-U finally converges to $2.6691 \times 10^{-5}$.
The loss function of MHP-V on the cross-validation set eventually converges to $3.8955 \times 10^{-6}$.

\subsubsection{Flow field prediction results}

\begin{figure*}[!h]
	\begin{center}
		\includegraphics[width=1 \linewidth]{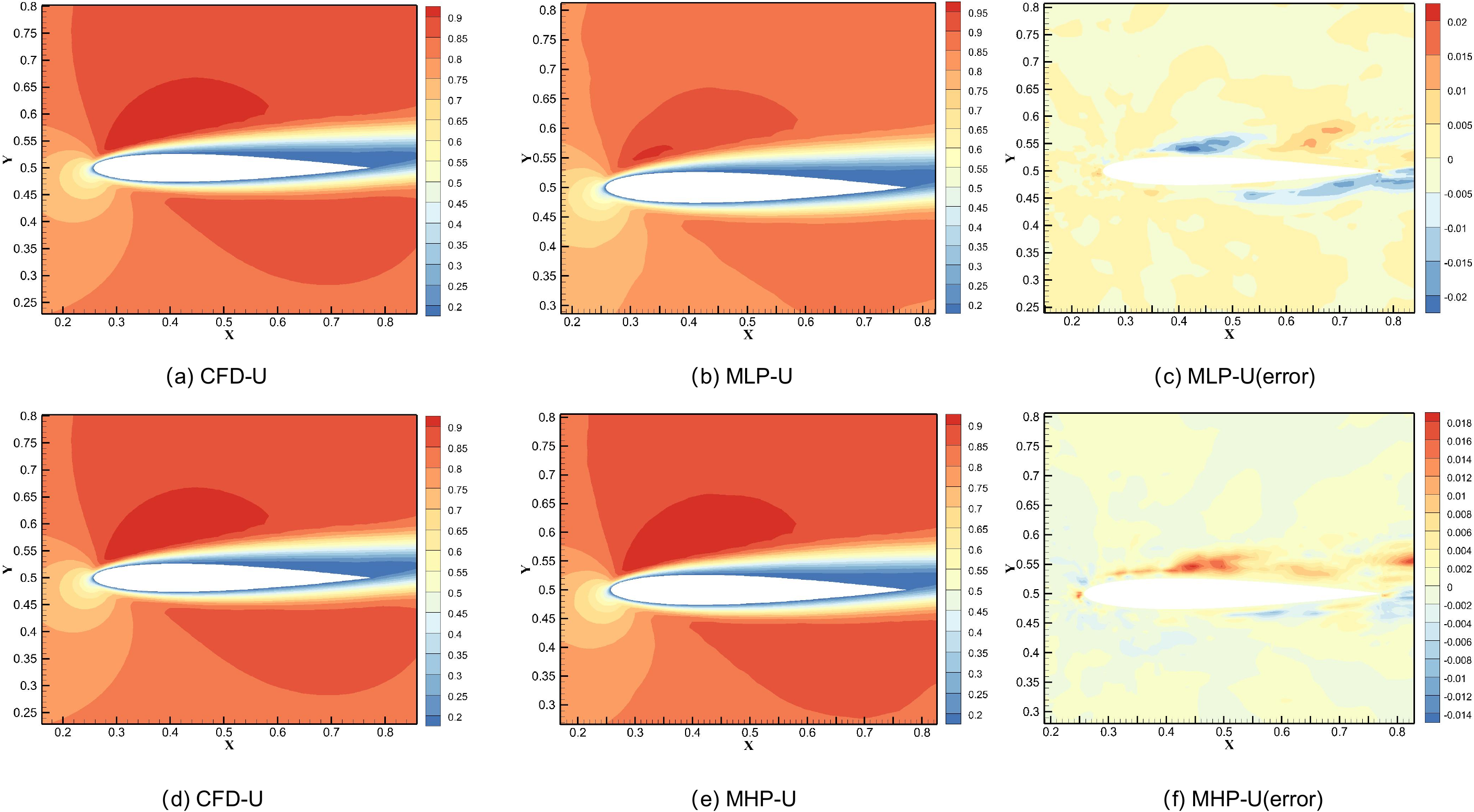}
	\end{center}  \vspace{-2mm}  
	\caption{{{Comparison of MHP-U, MLP and CFD calculation results. (a), (d) CFD calculation results for u-velocity. (b) U-velocity prediction results of MLP. (e) MHP-U prediction results for u-velocity. (c) Absolute error for u-velocity between CFD and MLP. (f) Absolute error for u-velocity between CFD and MHP-U.}
	}} \label{MLP_MHP_U} 
\end{figure*}

\begin{figure*}[!h]
	\begin{center}
		\includegraphics[width=0.9 \linewidth]{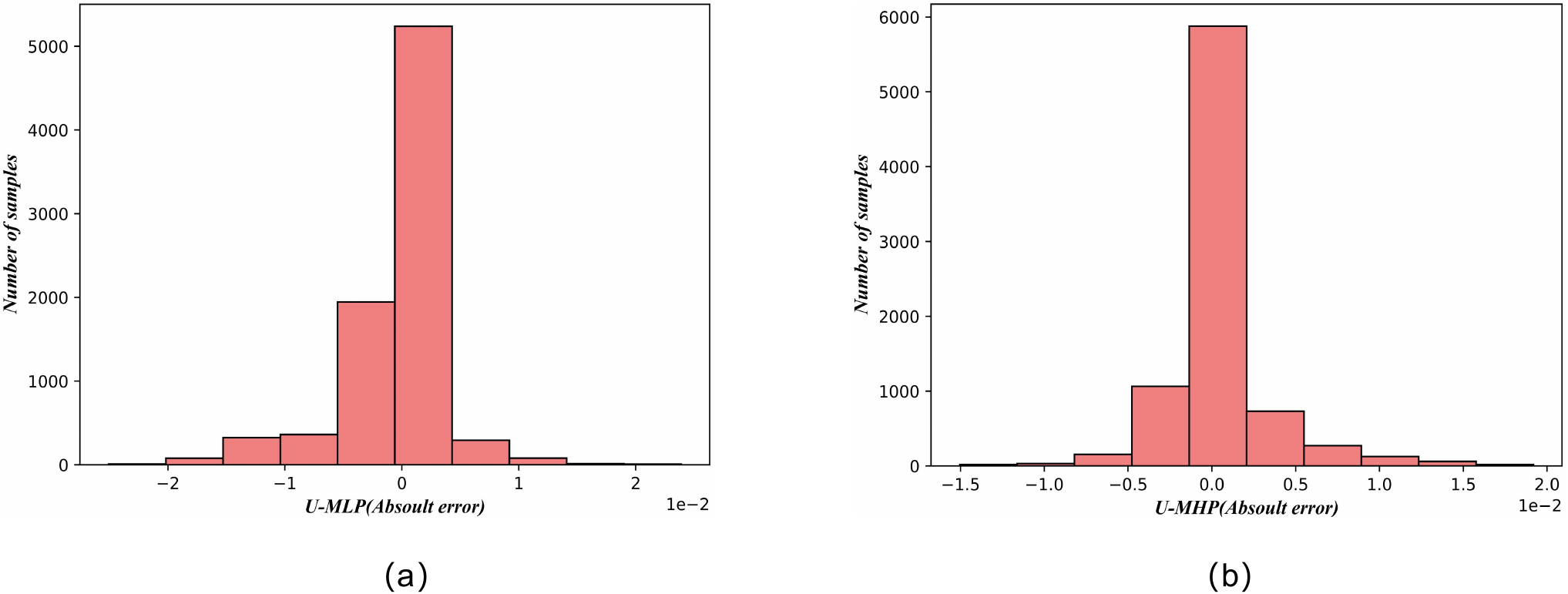}
	\end{center}  \vspace{-2mm}  
	\caption{{{Absolute error histogram between CFD and neural network. (a) Absolute error histogram of Fig. \ref{MLP_MHP_U}(c). (b) Absolute error histogram of Fig. \ref{MLP_MHP_U}(f)}
	}} \label{MLP_MHP_U_RES} 
\end{figure*}

Test the flow field prediction effect of MLP and MHP by selecting test data that the model has never seen before. 
In this research, NACA0012 airfoil at Re=1000 and AOA=6\degree \, is randomly selected to test the prediction capability of MLP and MHP. 
In Fig. \ref{MLP_MHP_U}, it can be found that the prediction results obtained by both MLP and MHP-U are consistent with the CFD calculation results. 
In addition, the absolute error plot between CFD and MLP, MHP-U is also given.
Figure. \ref{MLP_MHP_U}(c) shows that the absolute error range between MLP and CFD is -0.02 to 0.002. 
In Fig. \ref{MLP_MHP_U}(f), the absolute error range between CFD and MHP-U is -0.014 to 0.018.
Figure \ref{MLP_MHP_U_RES} uses the form of the histogram to show the error data distribution of Fig. \ref{MLP_MHP_U}(c) and Fig. \ref{MLP_MHP_U}(f). 
In Fig. \ref{MLP_MHP_U_RES}(a), about 5000 error data are distributed around 0. 
Figure \ref{MLP_MHP_U_RES}(b) there are about 6000 error data distributed in the numerical range of 0.
It can be seen from the test results that both MLP and MHP-U have a good prediction effect of u-velocity.
%
\begin{figure*}[!h]
	\begin{center}
		\includegraphics[width=1 \linewidth]{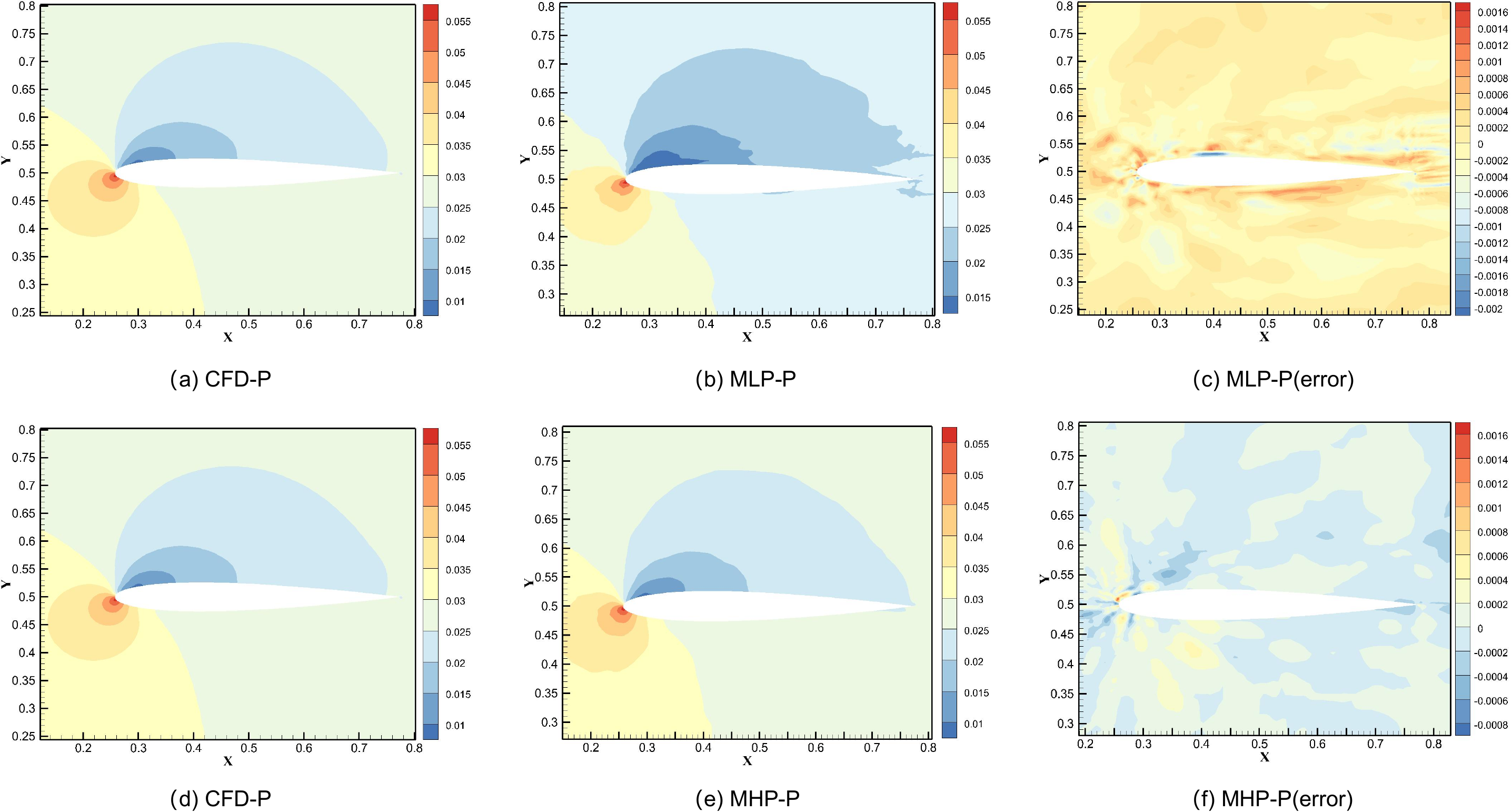}
	\end{center}  \vspace{-2mm}  
	\caption{{{Comparison of MHP-P, MLP and CFD calculation results. (a), (d) CFD calculation results for pressure. (b) Pressure prediction results of MLP. (e) MHP-P prediction results for pressure. (c) Absolute error for pressure between CFD and MLP. (e) Absolute error for pressure between CFD and MHP-P.}
	}} \label{MLP_MHP_P} 
\end{figure*}
\begin{figure*}[!h]
	\begin{center}
		\includegraphics[width=0.9 \linewidth]{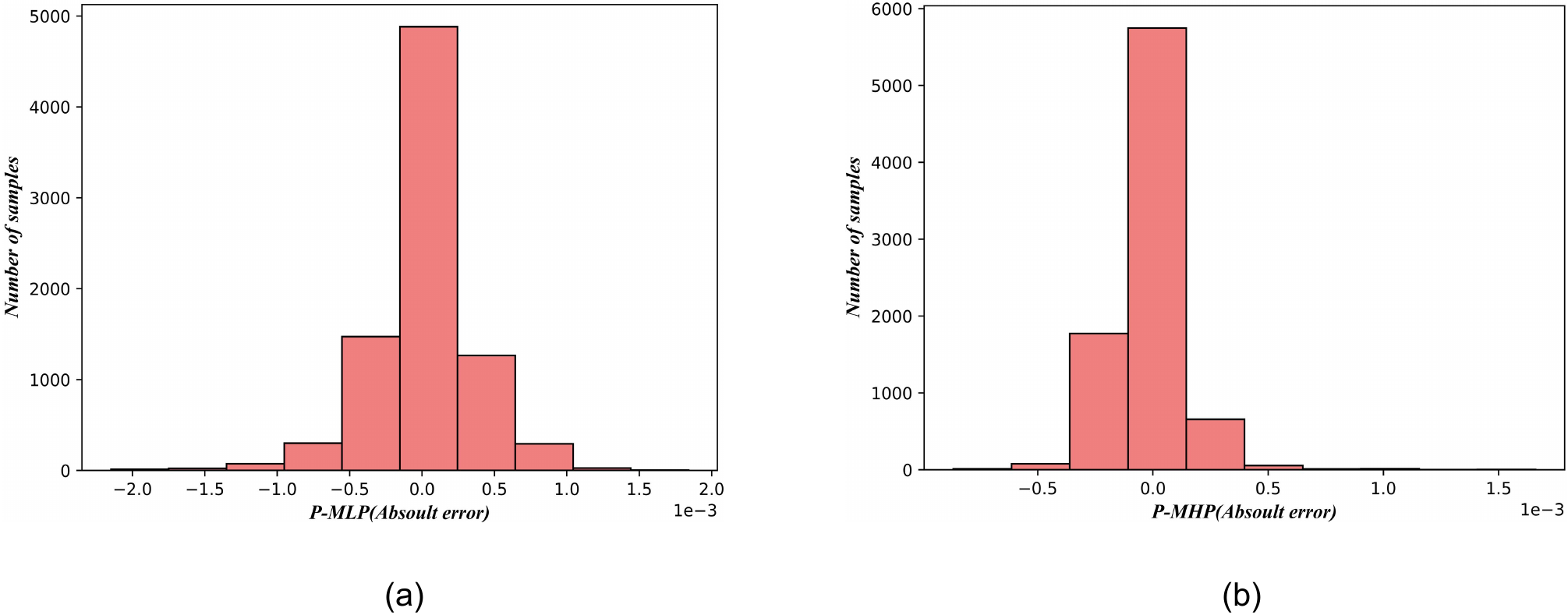}
	\end{center}  \vspace{-2mm}  
	\caption{{{Absolute error histogram between CFD and neural networks. (a) Absolute error histogram of Fig. \ref{MLP_MHP_P}(c). (b) Absolute error histogram of Fig. \ref{MLP_MHP_P}(f).}
	}} \label{MLP_MHP_P_RES} 
\end{figure*}
\begin{figure*}[!h]
	\begin{center}
		\includegraphics[width=1 \linewidth]{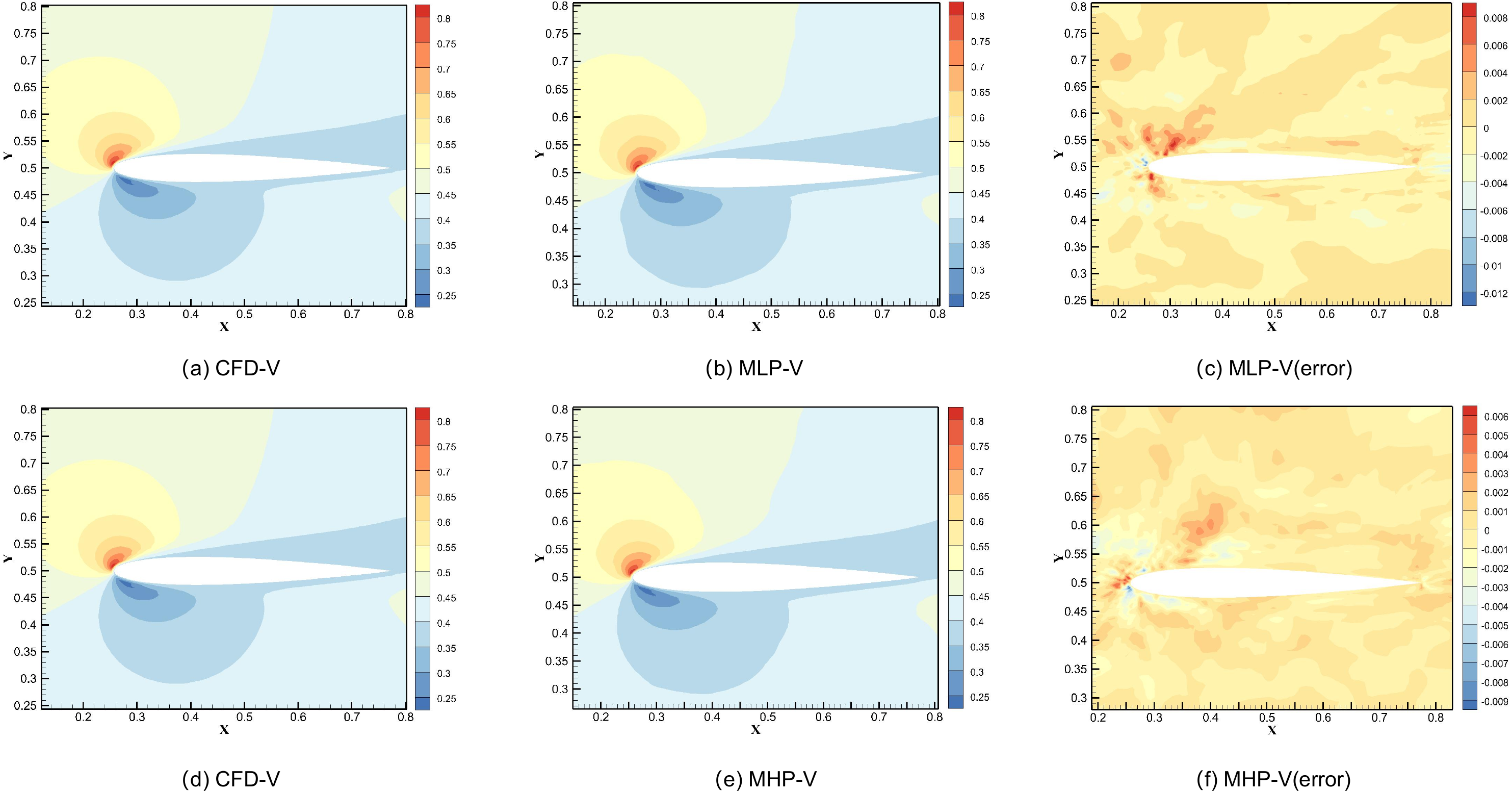}
	\end{center}  \vspace{-2mm}  
	\caption{{{Comparison of MHP-V, MLP and CFD calculation results. (a), (d) CFD calculation results for v-velocity. (b) MLP prediction results for v-velocity. (e) V-velocity prediction results of MHP-V. (c) Absolute error for v-velocity between CFD and MLP. (f) Absolute error for v-velocity between CFD and MHP-V.}
	}} \label{MLP_MHP_V} 
\end{figure*}
\begin{figure*}[!h]
	\begin{center}
		\includegraphics[width=0.9 \linewidth]{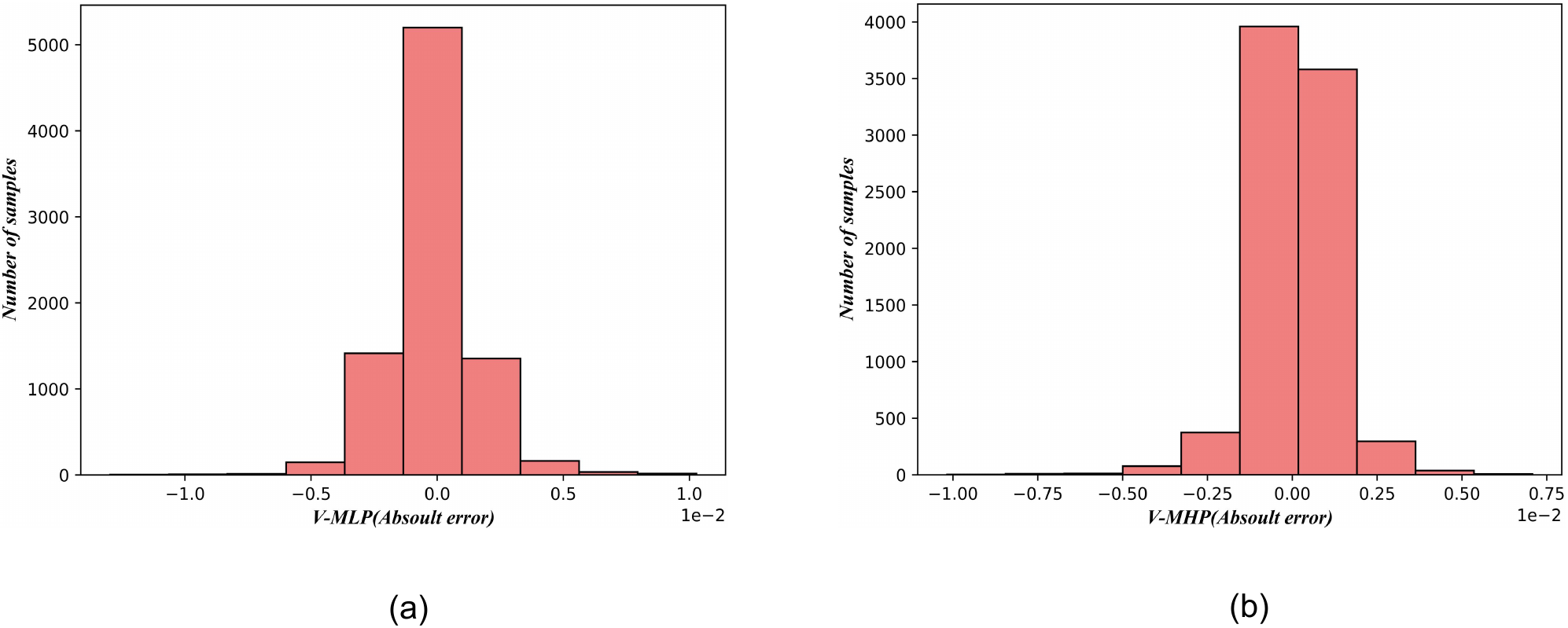}
	\end{center}  \vspace{-2mm}  
	\caption{{{Absolute error histogram between CFD and neural networks. (a) Absolute error histogram of Fig. \ref{MLP_MHP_V}(c). (b)Absolute error histogram of Fig. \ref{MLP_MHP_V}(f).}
	}} \label{MLP_MHP_V_RES} 
\end{figure*}
\begin{figure*}[!h]
	\begin{center}
		\includegraphics[width=1 \linewidth]{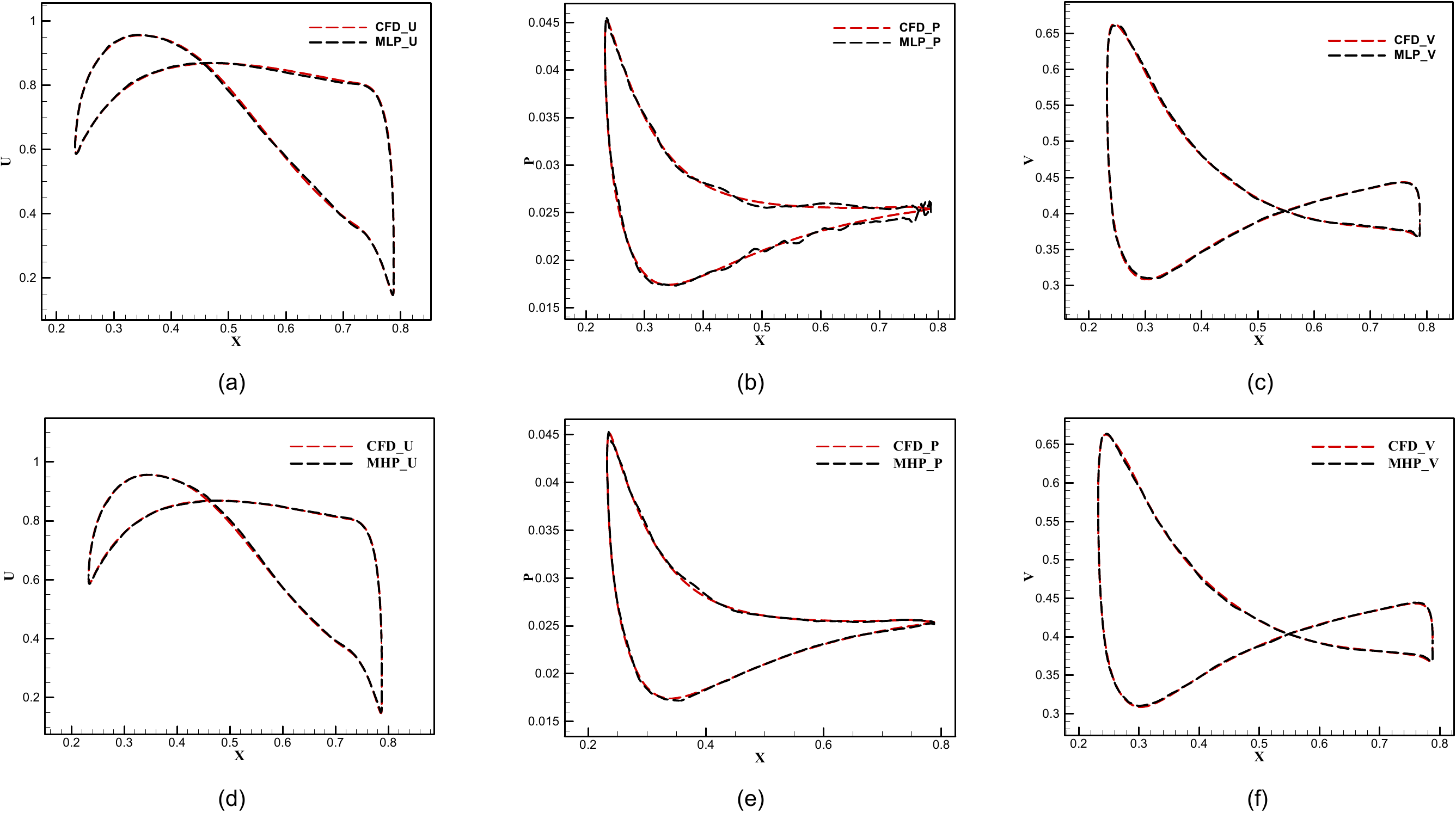}
	\end{center}  \vspace{-2mm}  
	\caption{{{Comparison of variables distribution between CFD and MLP, MHP. (a) Distribution of CFD and MLP about variable U. (b) Distribution of CFD and MLP about variable P. (c) Distribution of CFD and MLP about variable V. (d) Distribution of CFD and MHP-U about variable U. (e) Distribution of CFD and MHP-P about variable P. (f) Distribution of CFD and MHP-V about variable V.}
	}} \label{near-wall} 
\end{figure*}
\begin{figure*}[!h]
	\begin{center}
		\includegraphics[width=1 \linewidth]{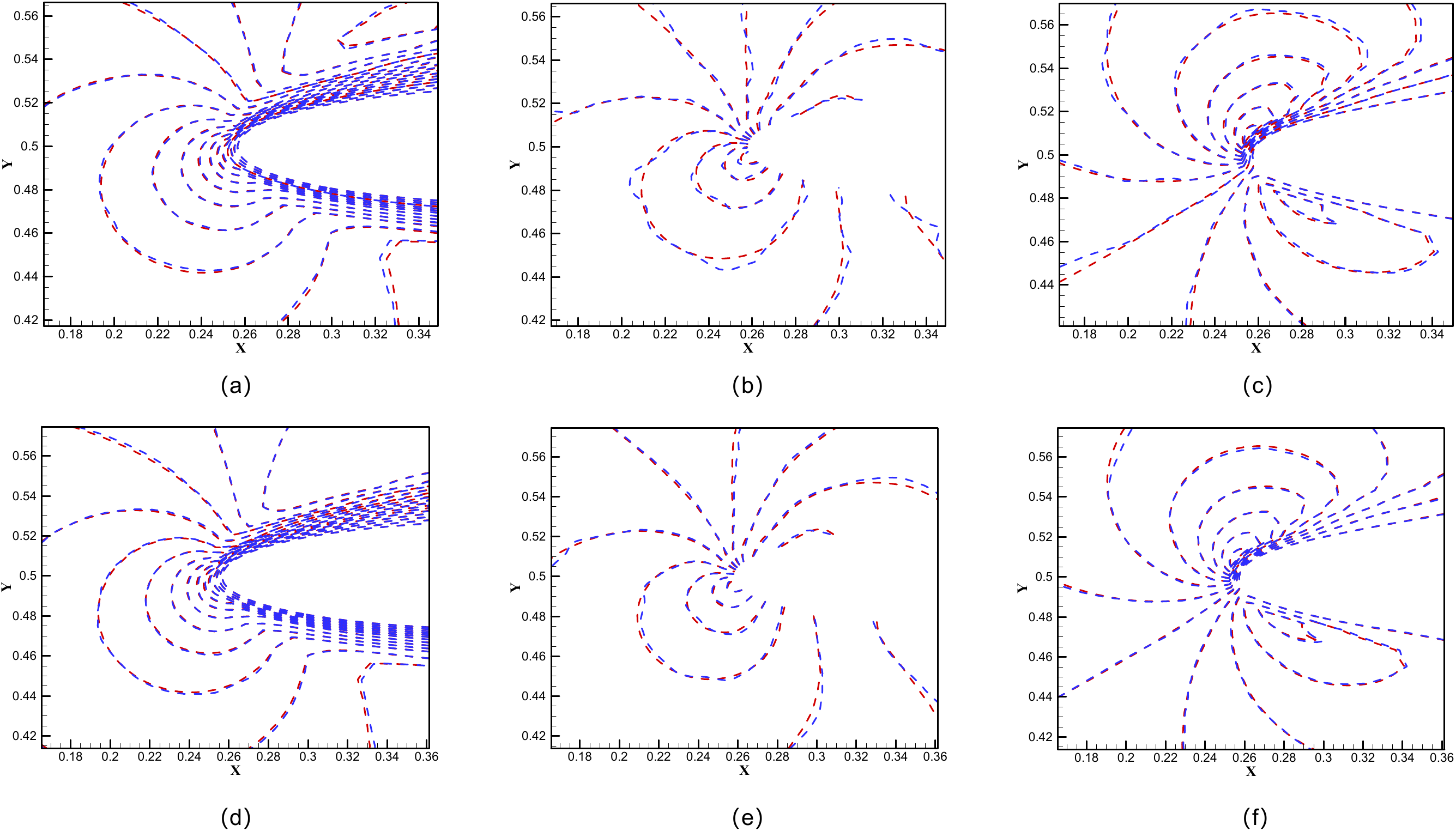}
	\end{center}  \vspace{-2mm}  
	\caption{{{Comparison of local contour of CFD and MLP, MHP. The red dotted line is CFD, the blue dotted line is neural network. (a) U-velocity contour of MLP and CFD. (b) Pressure contour of CFD and MLP. (c) V-velocity contour of CFD and MLP. (d) U-velocity contour of CFD and MHP-U. (e) Pressure contour of CFD and MHP-P. (f) V-velocity contour of CFD and MHP-V.}
	}} \label{MLP_MHP equal value} 
\end{figure*}
\begin{figure*}[!h]
	\begin{center}
		\includegraphics[width=1 \linewidth]{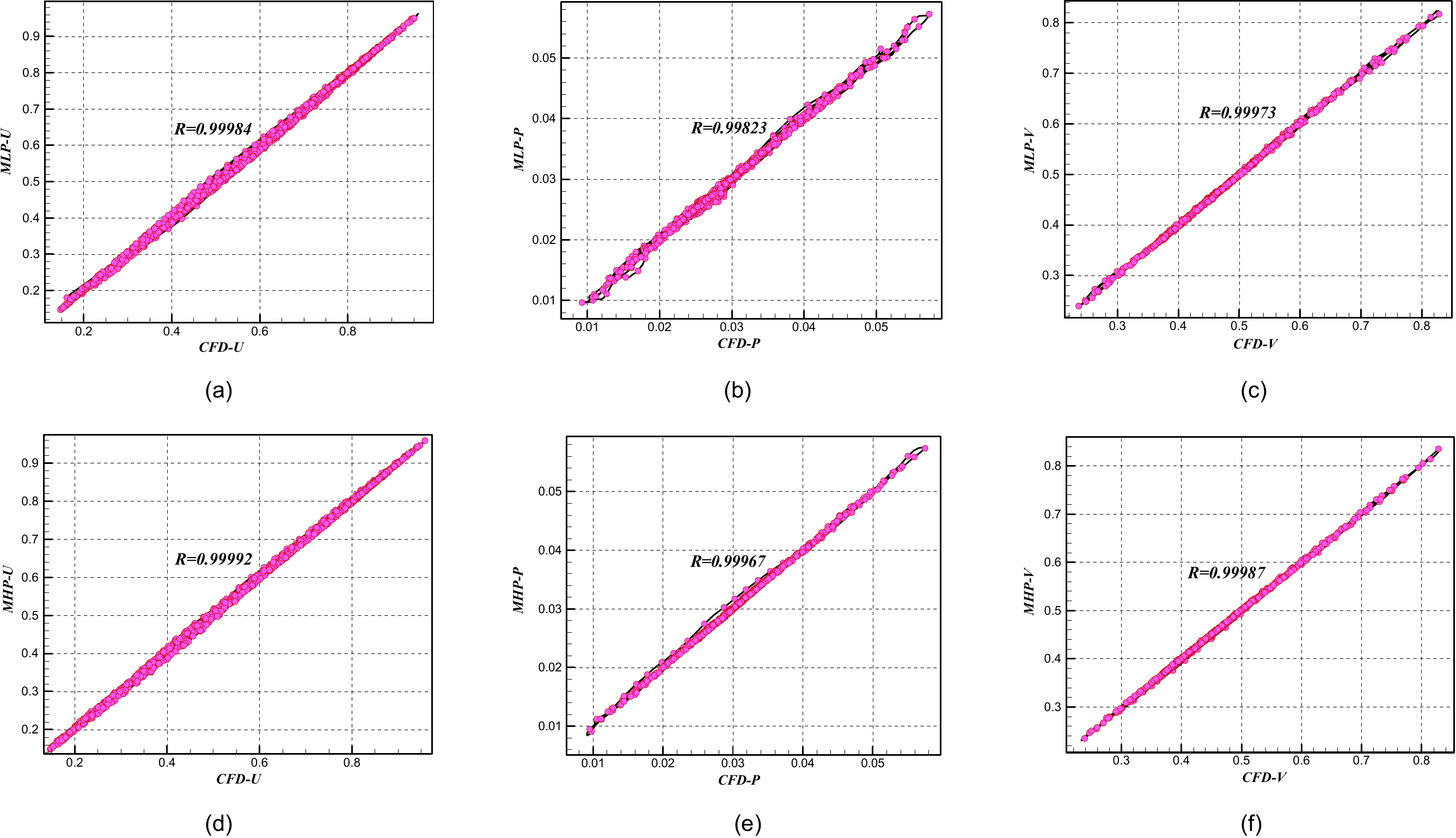}
	\end{center}  \vspace{-2mm}  
	\caption{{{Correlation curve of flow field prediction variables. (a) Correlation curve of CFD and MLP. (b) CFD and MLP correlation curve. (c) Correlation curve of CFD and MLP. (d) CFD and MHP-U correlation curve. (e) Correlation curve of CFD and MHP-P. (f) CFD and MHP-V correlation curve.}
	}} \label{relative_coefficitive} 
\end{figure*}
In Fig. \ref{MLP_MHP_P}, it can be found that the pressure prediction results of MHP-P is consistent with the CFD calculation results.
But the error between MLP prediction results and CFD calculation results is relatively large. 
And it can also be seen from the error distribution plot of Fig. \ref{MLP_MHP_P}(c) and Fig. \ref{MLP_MHP_P}(f). 
The error range between prediction values and ground truth of MLP is between -0.002 and 0.0016. 
And the error range between prediction values and the ground truth of MHP-P is between -0.0008 and 0.0016.
Figure \ref{MLP_MHP_P_RES} uses the form of histogram to show the error data distribution of Fig. \ref{MLP_MHP_P}(c) and Fig. \ref{MLP_MHP_P}(f).
And it can be clearly seen from the histogram that prediction error of MHP-P is more concentrated in the numerical range of 0, while the prediction error of MLP is more dispersed.
According to the pressure prediction results, for sparse data, MHP-P achieves better prediction effect than MLP.
In Fig. \ref{MLP_MHP_V}, both MLP and MHP-V achieve a good prediction results for v-velocity.
As can be seen from Fig. \ref{MLP_MHP_V}(c), the error between prediction results of MLP and CFD calculation results is between -0.012 and 0.008. 
In Fig. \ref{MLP_MHP_V}(f), the error between MHP-V prediction results and CFD calculation results is between -0.009 and 0.006. 
Figure \ref{MLP_MHP_V_RES} uses the form of histogram to show the error data distribution of Fig. \ref{MLP_MHP_V}(c) and Fig. \ref{MLP_MHP_V}(f).
And it can be found from Fig. \ref{MLP_MHP_V_RES}(a) that for MLP prediction error data, there are about 5000 error data distributed in the numerical range of 0.
And for the MHP-V prediction error data, there are about 7000 error data distributed in the numerical range of 0.
This shows that under the same conditions, the prediction effect of MHP-V is better than MLP.

From Fig. \ref{near-wall}, in the near wall region of the airfoil, the prediction effect of MLP and MHP on the u-velocity and v-velocity is relatively good.
This is because the velocity distribution of the airfoil is relatively continuous, and the neural network can quickly learn the relevant distribution law of the data during training process. 
However, it can be found from Fig. \ref{near-wall}(b) that due to the uneven distribution of pressure data, the curve obtained by MLP when performing the task of pressure prediction is not smooth, and the fitting effect with CFD is poor.
On the contrary, the pressure curve predicted by MHP is smoother.
The experimental results in Fig. \ref{near-wall}(e) shows that MHP-P still achieve a good prediction results in the face of sparse flow field data. 
It shows that the network architecture of MHP has better generalization than MLP.      

In Fig. \ref{MLP_MHP equal value}, for contours of u-velocity and v-velocity, both MLP and MHP have achieved a good prediction effect.
But due to the existence of sparse data, the prediction effect of MLP is slightly worse than that of MHP.
In particular, it can be found from Fig. \ref{MLP_MHP equal value}(b) and Fig. \ref{MLP_MHP equal value}(e) that this comparison is more obvious.
The pressure contours predicted by MLP is not smooth, resulting in poor fitting effect with CFD contours.
On the contrary, for sparse pressure data, MHP network architecture is decoupled, so the influence of sparse data on other parameters is avoided.
Moreover, MHP can focus more attention on sparse data, which makes the prediction effect of the model better and the generalization performance stronger in the face of sparse data. 
Figure \ref{MLP_MHP equal value}(e) shows that the contours predicted by MHP-P is consistent with the CFD calculation results.

The accuracy of the model is further verified by using the correlation coefficient between the ground truth and prediction values of the flow field. 
The correlation coefficient is calculated in the similar way to Equation (9), except that the parameters $T$ and $P$ here represents the ground truth and prediction values of the flow field, respectively.
In Fig. \ref{relative_coefficitive}(a), for u-velocity, the correlation coefficient between the prediction values and ground truth of MLP is $R=0.99984$. 
From Fig. \ref{relative_coefficitive}(b), for $C_p$, the correlation coefficient between the prediction values and ground truth of MLP is $R=0.99823$.
Figure \ref{relative_coefficitive}(c) shows that for v-velocity, the correlation coefficient between the prediction values and the ground truth of MLP is $R=0.99973$.
From Fig. \ref{relative_coefficitive}(d), Fig. \ref{relative_coefficitive}(e) and Fig. \ref{relative_coefficitive}(f), it can be found that the correlation coefficients between the prediction values and the ground truth of MHP for u-velocity, pressure and v-velocity are $R=0.99992$,  $R=0.99967$ and $R=0.99987$, respectively.
Secondly, it can be seen from Fig. \ref{relative_coefficitive} that most of the discrete points are distributed around the diagonal of image, indicating that the difference between ground truth and prediction values is small.
However, it can be found from Fig. \ref{relative_coefficitive}(b) that the correlation data points between the prediction values and the ground truth of MLP for pressure are scattered near the image diagonal, indicating that the prediction results of MLP for pressure is not accurate.
And for the same data, as shown in Fig. \ref{relative_coefficitive}(e), the correlation data between the prediction values and the ground truth of MHP-P are more closely distributed in the diagonal area of the image, indicating that the neural network architecture of MHP can obtained better prediction results even for sparse flow field data.

\section{Conclusions}

CNN is used to establish the mapping relationship between airfoil geometry shape and airfoil coordinates. 
Firstly, the input airfoil image is encoded into sixteen geometric parameters by CNN.
And then uses a decoder-like network architecture to decode these geometric parameters into the $y$ coordinates of corresponding airfoil image. 
The prediction capability of CNN network is tested on the test set of airfoil.
In the case of variable geometry, the correlation coefficient R=0.9999 between prediction values and ground truth of airfoils. 
Compared with the traditional airfoil parameterization methods, the deep learning method is more flexible.
Moreover, based on the pre-training model, the training set can be further expanded to improve the generalization of CNN.

For sparse flow field data, multi-head perceptron neural network architecture is proposed to improve the accuracy and generalization of flow field prediction. 
Firstly, the influence of the number of neural network layers and the number of neurons on the prediction accuracy of flow field is verified.
According to the experimental results, the network architecture of 10 layers and 180 neurons in each layer is selected as the basic network for MLP and MHP. 
After experiment comparison, it can be found that for sparse flow field data, MHP can achieve better prediction results than MLP.
This is because in the multi-variable prediction task, the sparse flow field data will cause the neural network to pay insufficient attention to it in the training process due to the lack of data. 
Moreover, in the multi-coupling architecture of MLP, multiple variables will interfere with each other in the process of neural network parameter updating.
Therefore, MHP decouples the multi-variable prediction task of flow field to avoid the interference of sparse data to the prediction results of other normal flow field data.
The loss function of MHP in training set and validation set can be reduced by two orders of magnitude compared with MLP.
And in order to test the prediction effect of MHP and MLP for different airfoil flow fields, Appendix A presents the flow field prediction results of MLP and MHP for NACA0024 at Re=1000 and AOA=8\degree.

\section *{Declaration of competing interest}
The authors declare that they are have no known competing financial interests or personal relationships that could have appeared to influence the work reported in this paper.

\section *{} 

\bibliography{mybibfile}

\newpage
\setcounter{figure}{0} 
\appendix 

\section{NACA0024 FLOW FIELD PREDICTION RESULTS}

\begin{figure*}[!h]
	\begin{center}
		\includegraphics[width=1 \linewidth]{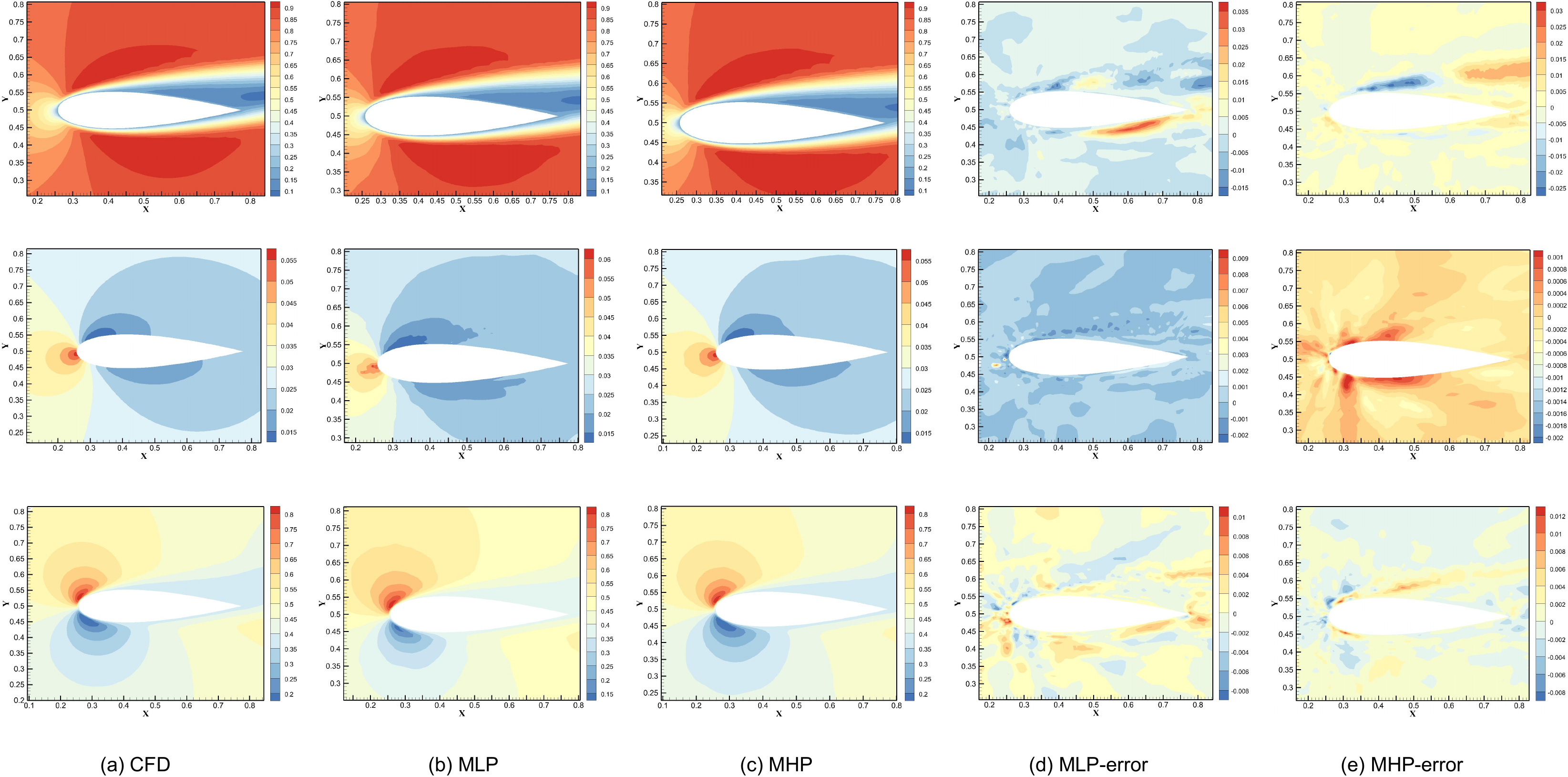}
	\end{center}  \vspace{-2mm}  
	\caption{{{Comparison diagram of CFD, MLP and MHP calculation results for NACA0024. First column: CFD calculation results. Second column: Flow field prediction results of MLP. Third column: MHP flow field prediction results. Fourth column: Absolute error diagram between MLP and CFD. Fifth column: Absolute error diagram between CFD and MHP.}
	}} \label{naca0024_upv} 
\end{figure*}
\begin{figure*}[!h]
	\begin{center}
		\includegraphics[width=1 \linewidth]{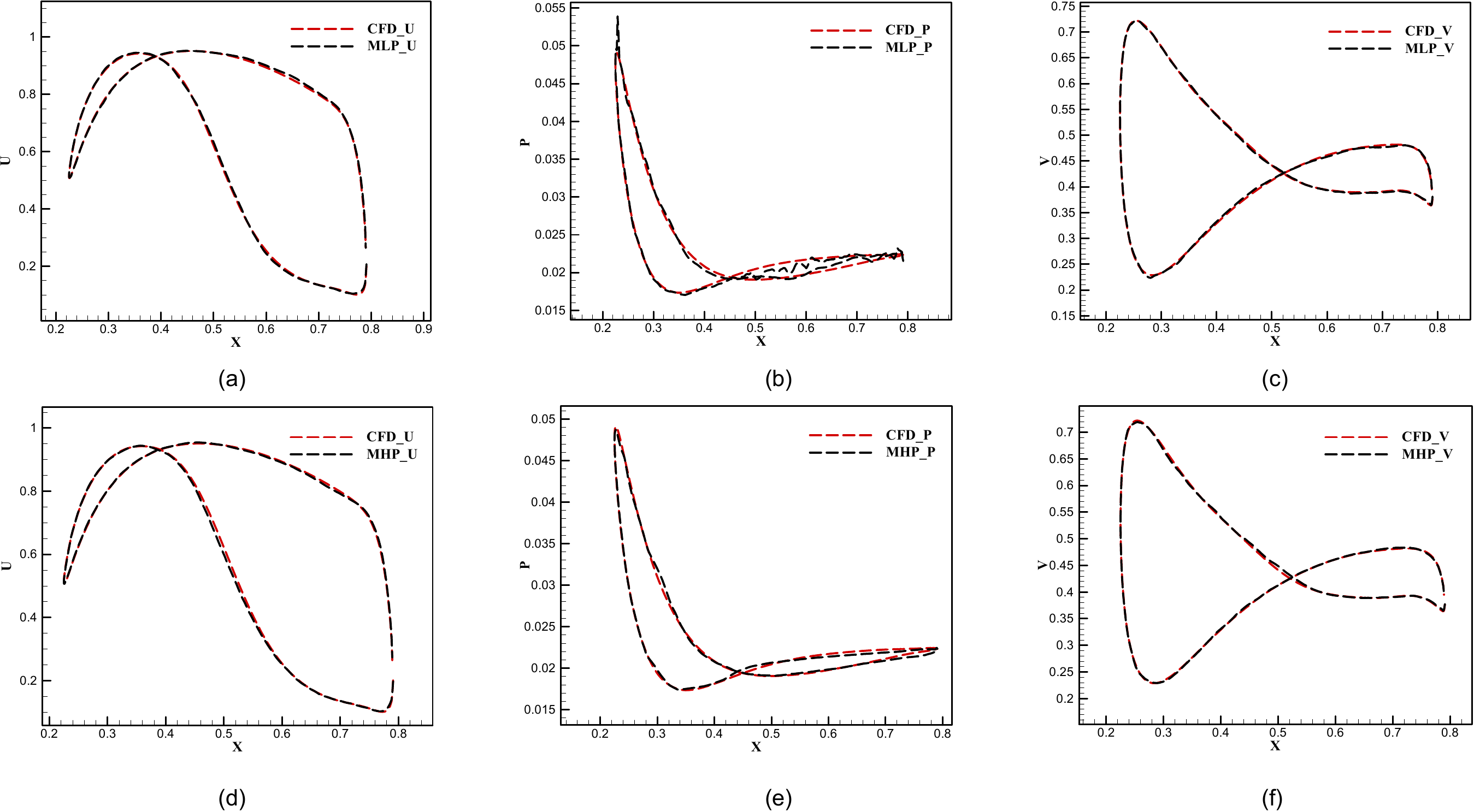}
	\end{center}  \vspace{-2mm}  
	\caption{{{Comparison diagram of variables distribution between CFD and MLP, MHP. (a) Distribution of CFD and MLP about variable U. (b) Distribution of CFD and MLP about variable P. (c) Distribution of CFD and MLP about variable V. (d) Distribution of CFD and MHP-U about variable U. (e) Distribution of CFD and MHP-P about variable P. (f) Distribution of CFD and MHP-V about variable V}
	}} \label{naca0024_near_wall} 
\end{figure*}
\begin{figure*}[!h]
	\begin{center}
		\includegraphics[width=1 \linewidth]{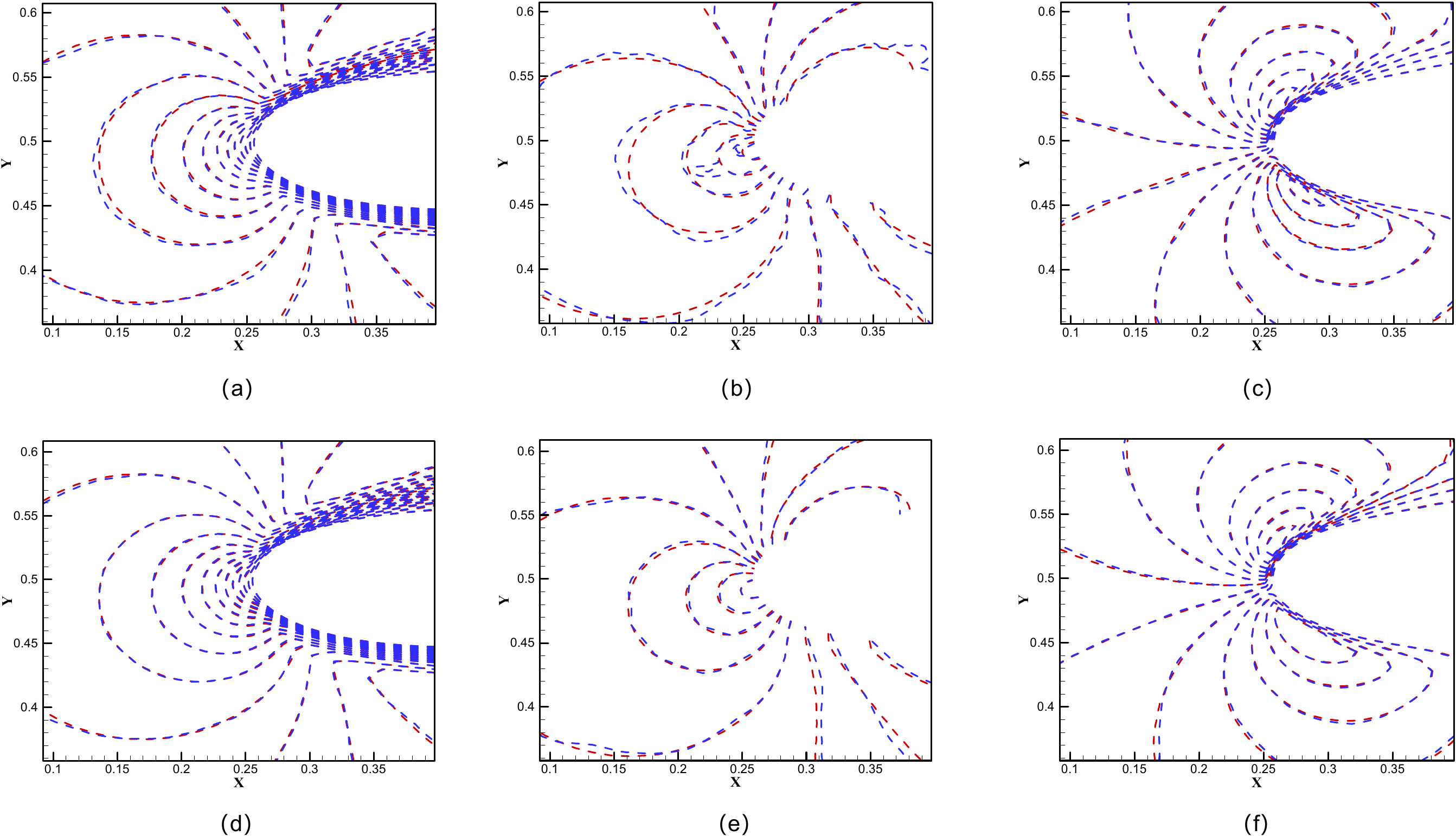}
	\end{center}  \vspace{-2mm}  
	\caption{{{Comparison diagram of local contour of CFD and MLP, MHP. The red dotted line is CFD, the blue dotted line is neural network. (a) U-velocity contour of MLP and CFD. (b) Pressure contour of CFD and MLP. (c) V-velocity contour of CFD and MLP. (d) U-velocity contour of CFD and MHP-U. (e) Pressure contour of CFD and MHP-P. (f) V-velocity contour of CFD and MHP-V.}
	}} \label{naca0024_equal_line} 
\end{figure*}
\begin{figure*}[!h]
	\begin{center}
		\includegraphics[width=1 \linewidth]{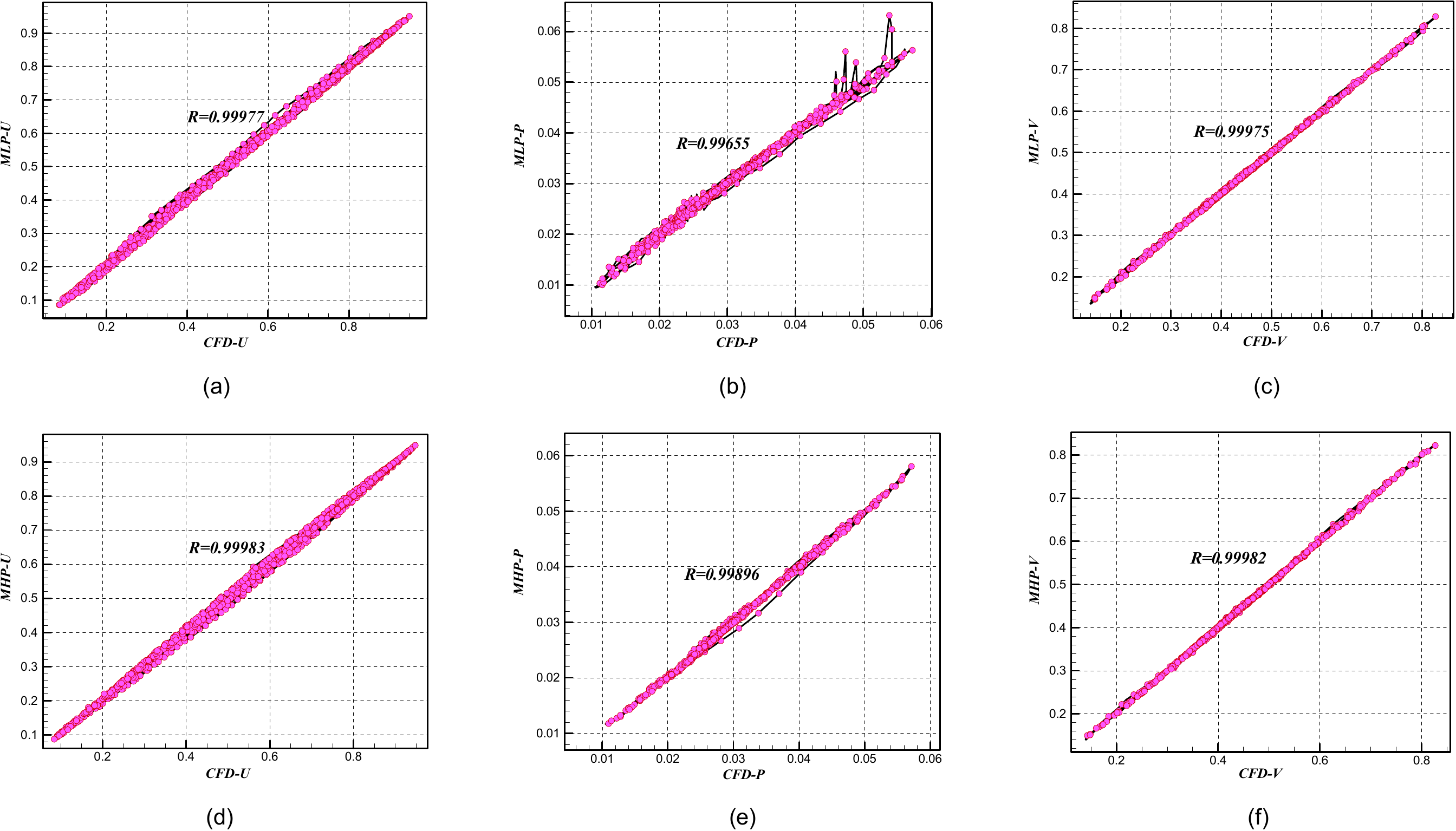}
	\end{center}  \vspace{-2mm}  
	\caption{{{Correlation curve of flow field prediction variables. (a) Correlation curve of CFD and MLP. (b) CFD and MLP correlation curve. (c) Correlation curve of CFD and MLP. (d) CFD and MHP-U correlation curve. (e) Correlation curve of CFD and MHP-P. (f) CFD and MHP-V correlation curve.}
	}} \label{naca0024_co} 
\end{figure*}
\end{document}